\documentclass{sig-alternate}
\usepackage[german,english]{babel}      
\usepackage[latin1]{inputenc} 
\usepackage{subfigure}    
\usepackage{amsmath}        
\usepackage{amssymb}        
\usepackage{dsfont}        
\usepackage{mathrsfs}
\usepackage{array}
\usepackage{mdwmath}
\usepackage{wrapfig}
\usepackage{url}
\usepackage{breakcites}
\usepackage{placeins}
\usepackage{color}
\usepackage[breaklinks]{hyperref}
\hypersetup{
    colorlinks=true,       
    linkcolor=black,          
    citecolor=black,        
    filecolor=black,      
    urlcolor=black           
}
\usepackage{multirow}    
\begin{document}
\newcommand{\emphtwo}[1]{\textit{#1}}
\title{Spatio-Temporal Small Worlds for Decentralized Information Retrieval in Social Networking}
\numberofauthors{3} 
\author{
\alignauthor
Georg Groh\\
       \affaddr{TU München}\\
       \affaddr{Faculty for Informatics}\\
       \email{grohg@in.tum.de}
\alignauthor
Florian Straub\\
       \affaddr{ETH Zürich}\\
       \affaddr{Inst. of Cartography and Geoinformation}\\
        \email{straubf@ethz.ch}
\alignauthor 
Benjamin Koster\\
       \affaddr{TU München}\\
       \affaddr{Faculty for Informatics}\\
       \email{koster@in.tum.de}
}

\maketitle
\begin{abstract}
We discuss foundations and options for alternative, agent-based information retrieval (IR) approaches in Social Networking, especially Decentralized and Mobile Social Networking scenarios. In addition to usual semantic contexts, these approaches make use of long-term social and spatio-temporal contexts in order to satisfy conscious as well as unconscious information needs according to Human IR heuristics. Using a large Twitter dataset, we investigate these approaches and especially investigate the question in how far spatio-temporal contexts can act as a conceptual bracket implicating social and semantic cohesion, giving rise to the concept of Spatio-Temporal Small Worlds. 
\end{abstract}
%
%
%
\section{Introduction}
Social Networking (SN) and Decentralized Social Networking (DSN) \cite{yeung2009decentralization} as a future variant of SN is extensively used to build rich personal and interpersonal information spaces. Furthermore, the increased access of SN-platforms via mobile devices such as smartphones (giving rise to new paradigms such as (context-aware) Mobile Social Networking (MSN)) introduces a steeply growing permeation  of these information spaces with explicit spatial context. Thus, besides social contexts such as `friendship' relations, spatio-temporal contexts and their interrelations with social contexts are also available and extensively used in modern (M)SN platforms. 

These upcoming SN paradigms allow users more and more to employ special forms of information retrieval, akin to traditional human information seeking behavior based on the real social network of society  (`Human IR') which, besides semantic context, also uses social and spatio-temporal context (see also \cite{wilson2000human}). 

Inspired by this behavior, the question now arises how alternative IR services for SN may be constructed that effectively make use of social, semantic, and spatio-temporal contexts and their interrelations.

Pursuing this research question, the reminder of this paper is structured as follows. After a brief discussion of the relation between context and information needs, we address Human IR and wayfinding in social networks. We then introduce the concept of Spatio-Temporal Small Worlds for IR in Social Networking as well as a respective architecture based on personal information agents. The following main part of the paper empirically investigates the concept of Spatio-Temporal Small Worlds and the 
suitability of the principles guiding alternative IR processes inspired by Human IR, using social search, semantic search and spatio-temporal search and here especially the suitability of spatio-temporal embedding as a contextual bracket using a large Twitter dataset.

%
This paper is an extended version of the content of the paper \cite{grohstraubkoster}. Elements of this text also appear in the thesis \cite{grohhabil}. 
%
\section{Related Work and Fundamental Considerations}
\subsection{Context and Unconscious Information \\ Needs}
\label{infoneeds}
%
%
In \cite{mizzaro1998many} adequate characterizations of \emph{relevance in information retrieval} (IR) and especially \emph{qualifications of information needs} that a user of IR has in view of a `problematic situation'
\cite{belkin1982askA},  \cite{belkin1982askB},  \cite{mizzaro1998many} are investigated. In this regard, the concepts query, request, perceived information need (PIN), and real information need (RIN) are considered as central. The query is a formalization of a request which, in turn, is a natural language expression of a PIN. The PIN is the information need that a user subjectively \emphtwo{perceives} in the problematic situation. The RIN may e.g. be defined via the entirety of information that is `objectively' relevant for the solution of the problem, thus \emphtwo{extensionally defining} the `problem' in `problematic situation' \emphtwo{through} the RIN. `Objectively' may e.g. be determined by the intersection or union of the assessed RIN by the fictional set of all human experts for the problem. 
%

 During the IR process the user then consumes or partly consumes the results, uses his assessment of relevance judgments, corrects his PIN, formulates a new query and so on, giving rise to a \emph{circular IR process} (see e.g. \cite{belkin1993interaction}). 
A user will explore the space of information relevant to the RIN by repeated executions of the aforementioned IR cycle, iteratively re-shaping his PIN, and enlarging the set of acquired information. 

Our notion of \emphtwo{conscious information need} corresponds to \emphtwo{perceived information need} (PIN) in \cite{mizzaro1998many} and our notion of \emphtwo{unconscious information need} encompasses the \emphtwo{real information need} (RIN) in \cite{mizzaro1998many}. 
In IR, the term unconscious information need is justified because the user is not consciously aware of information needs in RIN $\setminus$ PIN (that are in RIN but not in PIN) in a `problematic situation'. However, our notion of \emphtwo{unconscious information need} also encompasses \emphtwo{an unspecific readiness to accept `interesting' information}. Unless artificially defining some `background problematic situations', \emph{ongoing readiness to accept welcomed information that does \emphtwo{not} correspond to a `problematic situation'} (and thus not to a RIN or PIN) is not represented in the schema of IR relevance. This case is simply not covered by the concept of \emphtwo{information retrieval}, where a problematic situation induces a concrete information need which in turn finally induces a query. Examples for such a form of \emphtwo{unconscious information need} correspond to e.g. a user reading `something interesting' on a news-feed or is being told `something interesting' by a friend, etc. Thus information may be delivered to a user that the user has no a priori \emphtwo{perceived} information need for, and which the user has not explicitly asked for via a query or filter, but that he / she nevertheless judges as `interesting'. This kind of information is usually pro-actively delivered by awareness services, or news services, or by direct communication services \cite{grohhabil}. 


\emph{Context} and especially \emph{social context} may be used to provide a \emph{relevance bracket} for this `interesting information' that is delivered to a user by such services by e.g. narrowing the visualizations of social network dynamics to the network neighborhood or spatio-temporal neighborhood of a user \cite{aaagrohLehmannWangHuberHammerl}, using social filtering to deliver horizon broadening recommendations \cite{aaagrohEhmig},  or using social contexts to specify suitable audiences for certain information \cite{aaagrohBirnkammerer}.

\begin{figure}[htb]
\centering
\includegraphics[width=0.95\columnwidth]{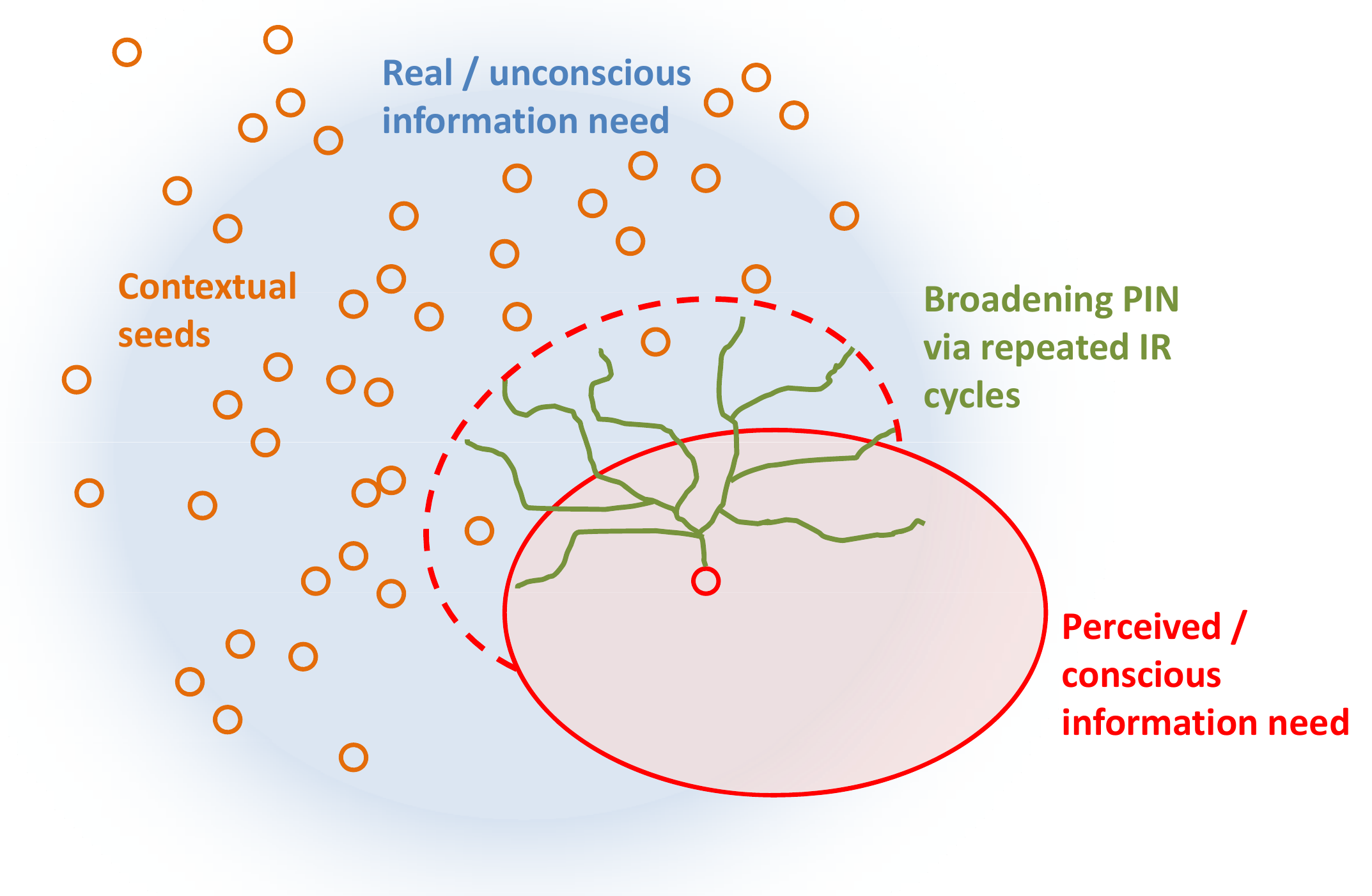}
\caption{Defining unconscious information need \cite{grohhabil}}
\label{figurecontextualIR}
\end{figure}

The \emph{contextual relevance bracket} is a means to anticipate or \emph{induce relevance via context} in these proactive services \cite{grohhabil}. \emph{Incorporation of context}, especially of social and spatio-temporal context, can \emph{be especially useful for information retrieval} e.g. by aiding the user in exploring the space of relevant information items / in expanding the PIN, especially in relation to problems for which the RIN is hard to determine. This aid can be achieved by seeding the IR cycle with new motives especially beyond the PIN while providing a certain contextual bracket for relevance (in contrast to e.g. randomly choosing the seeds) as \autoref{figurecontextualIR} illustrates. 
In contrast to well defined problems, which may exhibit a natural saturation effect in view of new information, insights, competence gains, or perspectives appearing after new IR cycles and thus PIN$\approx$RIN after `sufficiently' many IR cycles, the `problematic situations' for which the RIN is hard to determine might not exhibit this saturation effect, either because the problem's definition is not precise enough or because the space of information items relevant to the RIN is very large. 

Traditional \emph{Context-Sensitive Information Retrieval} is usually focused on using types of context such as query histories or implicit feedback on the results to a query (e.g. via click-analysis or eye-tracking) to improve relevance of the immediately retrieved results in view of a given query (see e.g. \cite{shen2005context}). However it is usually limited to the PIN expressed in the query, because more general contextual brackets (e.g. induced by social context) that would be able to deliver the contextual seeds mentioned before are missing or not regarded. 
E.g. including seeds from the information spaces of other competent people determined via (besides the query) also taking social context into consideration, may improve the exploration of the RIN, especially in those cases where the boundaries of the RIN are hard to determine precisely.
\subsection{Human IR and Wayfinding in Social Networks}
\label{humanir}
If \emph{long-term social contexts} in the form of \emph{social networks} are used to provide \emph{contextual brackets} for information retrieval services in SN / MSN, it is important to review the basic results of \emph{decentralized routing and searching} in these networks \cite{liben2010wayfinding}. 

In 1967, \emph{Milgram's experiment} \cite{milgram1967small} showed that decentralized routing in social networks is possible and that the path lengths involved were small 
Watts and Strogatz \cite{watts1998collective} were able to provide a network model for such \emph{Small World networks}, which did not only explain their short mean average path length 
but also their high clustering coefficient (the network theoretic measure for triadic closure), a crucial property of social networks. 
The \emph{Watts-Strogatz model} is based on a toroidally, regularly linked graph, where edges are randomly redirected with a certain probability (shortcuts). These constructive elements generate the local cluster structure and  short mean average path length. \cite{watts1998collective}. 

While such models were able to explain the basic \emphtwo{structure of social networks}, the actual explanation of the Milgram experiment, the question of how \emph{decentralized wayfinding or routing} \emphtwo{can actually be accomplished}, was investigated by \emph{Kleinberg} \cite{kleinberg2000small}. In his variant of the Small World model, starting from a regularly linked network on a grid, the random distant re-connections of a node $a$ to a node $b$ were established with a probability $d(a,b)^{-\alpha}$. He was able to show that for $\alpha$ corresponding to the dimension of the grid, a decentralized (local knowledge only) routing algorithm, always choosing the node located closest to the target node as the next node, is sufficient to produce `sufficiently' short expected delivery times, polynomial in $O(log(n))$, where $n$ is the number of nodes in the network. Refinements of this model in view of more realistic \emph{geographic distributions of friendship relations} on the earth's surface were investigated by \cite{libennowell2005geographic}, suggesting a different geographic connection probability distribution and empirically finding a different value for $\alpha$, but confirming that the simple greedy local routing algorithm still leads to efficient delivery. 
This confirms that for \emph{efficient decentralized geographic routing in social networks}, the nodes (actors) of the network need to be \emph{spatially embedded} (e.g. have a known center of life) and each forwarding actor needs to have a \emph{cognitive model of this spatio-temporal context}.

More generally, besides spatial proximity other types of contextual metrics such as other long-term social contexts (e.g. occupation or hobbies) may as well be chosen to select the next node. The greedy local social search will select as the next node the node closest to the target node according to the given metrics (see e.g. \cite{liben2010wayfinding}). 

Parallels exist between using general context information for decentralized routing 
and the way \emph{social information retrieval} is accomplished in \emph{human societies}, which in turn has obvious \emph{commonalities with SN / MSN}. In `\emph{Human IR}', a question formalizing a PIN is `routed' to persons which presumably dispose of the required information in their (not necessarily properly explicated) information spaces. The resulting routes need to be `socially resilient' enough (e.g. in the sense of Granovetter's strong ties \cite{granovetter1973strength}) to support the actors en-route agreeing to process the query and to support routing the retrieved information back to the questioner.
At the same time the routes must contain enough weak ties (in Granovetter's sense) to convey new information or provide 
access to otherwise hardly reachable parts of the network via weak tie shortcuts in the sense of \cite{watts1998collective} 
\cite{centola2007complex}. 

As reviewed in \cite{wilson2000human}, \emph{human information seeking behavior} often use context e.g. social context to determine actors who could be asked, especially if the problem situation and  the PIN is poorly defined (\cite{wilson2000human}). 
Actors facing an informational problem will, besides the PIN ($\widehat{=}$ WHAT), evaluate \emph{all types of contexts}, their interrelations and their relations to the PIN, in order to render their PIN more precise, expand their PIN towards the RIN and ultimately collect enough information to solve their problem (see \cite{aaagrohStraub} for a more elaborate discussion). 
For the discussion, types of contexts will be represented by other interrogative pronouns such as WHO (pointing to social context), WHERE and WHEN (pointing to spatio-temporal context).
\emphtwo{Vice versa}, the asked persons may also use contextual knowledge to select appropriate information for the questioner, which may also include information that is not strictly relevant to the query but relevant to the PIN or even RIN of the questioner. Thus relevance may also be induced by the asked actor \emph{via contextual knowledge}. As an example consider the question ``How do I search for certain terms while I browse a text-document with UNIX `more' ?''. As an \emphtwo{expert}, a person might answer ``Use the `/' character and enter the term''. As an \emphtwo{expert \upshape{and} friend} the answer may include ``Besides: use `less' instead of `more'! It has a number of advantages''. As an \emphtwo{expert and \upshape{close} friend} the answer may include ``Besides: I give You the advice to quit using UNIX! A Mac will suit Your needs and the needs of Your wife much better. It provides more comfortable means to view and search text-files while still retaining `less' and `more' if desired'', using social context and the questioner's individual context. 

In terms of \emph{long-term social context}, \emph{Human IR `uses'} the main \emph{characteristics of small world networks} to \emph{search} in the complex network of distributed information spaces and context-elements for the right information: actors are able to reach experts (and their information spaces) via short expected path lengths and the highly clustered structure ensures that each actor has a large number of routing options. 
Suitable interdependent contextual metrics (Semantic (WHAT), social (WHO) or spatio-temporal (WHERE + WHEN)) allow efficiently \emph{navigating the space}. 
%
%
%
\section{Spatio-Temporal Small Worlds for IR in Social Networking}
\label{irapproach}
The question now arises, how we can employ these considerations and the considerations of the preceding section to \emph{construct an alternative information retrieval service for SN / MSN}.  While the complex socio-psychological mechanics of amalgamating and evaluating the interdependencies of WHO $\leftrightarrow$ WHERE $\leftrightarrow$ WHAT $\leftrightarrow$ WHEN  
in Human IR in view of searching the distributed information spaces in a context sensitive way are too intricate to model directly, \emph{spatio-temporal embedding may act as a reference point} and a means to naturally encode these interdependencies between the various forms of context for a respective IT model.  
\subsection{Spatio-Temporal Small Worlds}
A \emph{social spatio-temporal small world} may be defined as a social network, where the actor-nodes are \emphtwo{spatio-temporally embedded} into space-time via their current center of life (compare previous section). The relations correspond to directed long-term social relations of various types. We have seen that spatial distance metrics (and via using a current time-frame thus also spatio-temporal distance metrics) allow efficient decentralized routing. We assume that spatio-temporal distance metrics can thus also serve as one key means for a successful search for information in the social spatio-temporal small world part of the complex network of distributed information spaces and context-elements described in the previous section. `Successful' implies that the information found is relevant in view of a user's RIN especially in those cases where the RIN is hard to demarcate (see discussion in \autoref{infoneeds}). Another argument for using spatio-temporal distance metrics as a means to naturally encode interdependencies between the various forms of context or other metrics is that the studies of Kleinberg \cite{kleinberg2000navigation} and Liben-Nowell \cite{libennowell2005geographic} imply that in a social spatio-temporal small world, \emph{spatio-temporal closeness is probabilistically correlated with social closeness}. 

A \emph{semantic spatio-temporal small world} may be defined as a network of information items (e.g. documents) that are \emphtwo{spatio-temporally embedded} into space-time via certain criteria. Either 
 the information item's meta-data contains an \emph{explicit spatio-temporal embedding} or \emph{implicit spatio-temporal embedding}, e.g. explicated spatially via geo-parsing (see e.g. \cite{larson1996geographic} \cite{jones2008geographical}) and geo-coding (see e.g. \cite{jones2008geographical}) of the found named entities (see e.g. \cite{nadeau2007survey}). A third case applies if the information item is spatio-temporally embedded in the \emph{same spatio-temporal location(s) as the actor} whose information space this item is associated with.

The \emph{first mode of edges} of this network are the links indicating semantic relatedness of the items (e.g. HTTP links). The corresponding network has small world properties \cite{jin2006small}. The \emph{second mode of edges} relates items, whose `owners' are linked in the social spatio-temporal small world, which also gives rise to a network with small world properties. 

As previously discussed, \emph{social closeness is probabilistically correlated with spatial (and implicitly spatio-temporal) closeness} \cite{liben2010wayfinding} \cite{scellato2011socio}.

Studies by Brent Hecht \cite{hecht2008geosr}, \cite{hecht2008mapping}, \cite{hecht2009terabytes}, \cite{hecht2010localness} and others (e.g. \cite{lieberman2009you}) imply that in a semantic spatio-temporal small world, \emph{spatio-temporal closeness is probabilistically correlated with semantic closeness} to a certain extend, which is also expressed as a statistical tendency in (so-called) Tobler's first law of Geography: ``everything is related to everything else, but near things are more related than distant things'' \cite{tobler1970computer}. 
\begin{figure}[htb]
\centering
\includegraphics[width=0.85\columnwidth]{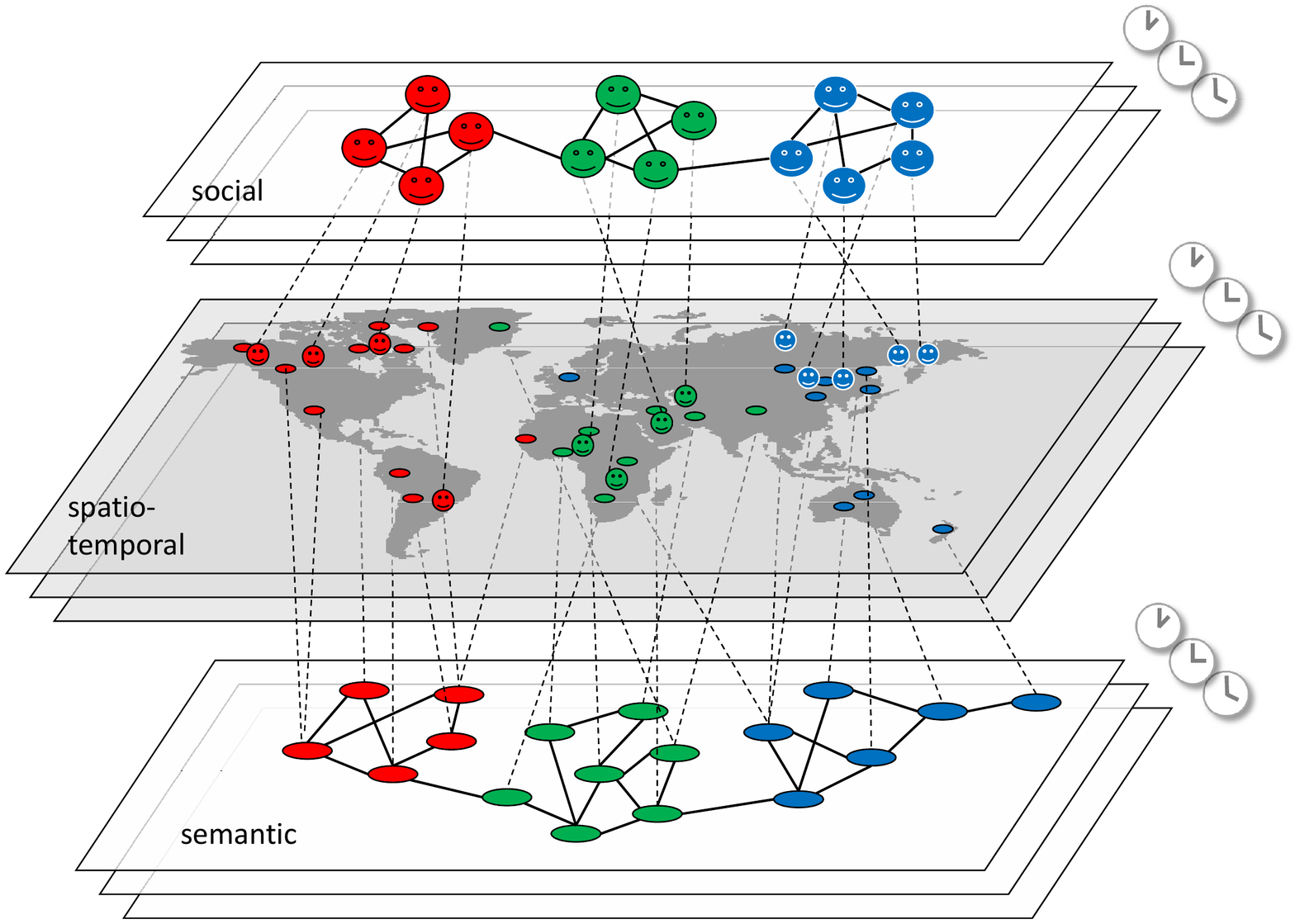}
\caption{
Spatio-temporal embedding of small worlds: how spatio-temporal embedding maintains social and semantic closeness properties as a statistical tendency \cite{grohhabil}\cite{aaagrohStraub}.
}
\label{figureStraub1}
\end{figure}

\autoref{figureStraub1} visualizes social and semantic spatio-temporal small worlds and illustrates the maintenance of social and semantic closeness via spatio-temporal embedding.

\emph{Social closeness is also probabilistically correlated with semantic closeness}. Homophily (the tendency of  similar people to associate with each other, contributing to triadic closure) \cite{mcpherson2001birds} and Peer Influence (the influence of persons which are directly linked in the social network) \cite{centola2010spread} can be prominently attributed for the local homogeneity in terms of information spaces of social groupings. The correlation between social closeness and semantic / topical closeness is also supported by other studies such as \cite{angelova2009investigating} and indirectly by \cite{aaagrohEhmig}.

Thus in view of decentralized search of relevant information in the complex network of distributed information spaces and context-elements which is characteristic of SN / MSN, we assume that social spatio-temporal small worlds and semantic spatio-temporal small worlds may act as a simple model of this complex network of context elements and spatio-temporal metrics may aid the decentralized search because of implicitly representing interrelations between spatio-temporal, social, and semantic relatedness. 
%
%

Based on these considerations and models and the principles of Human IR, the \emph{study} \cite{aaagrohStraub}, proposed a new \emph{context-aware, agent-based, federated approach to information retrieval in decentralized SN / MSN} in order to investigate limits and chances of using spatio-temporal embedding and its implicit `conservation' of semantic and social context as a contextual bracket. Besides the discussion of the last sections, the design decisions in this study were supported by a number of observations such as the ever growing availability of context in SN and especially MSN, the importance of the paradigm of Distributed Social Networking \cite{yeung2009decentralization}, the problems that the Hidden Web especially in connection with access protected SN / MSN information spaces generates for traditional search engines \cite{he2007accessing}, or the obvious parallels that searching in SN / MSN has to Human IR. 
\subsection{An Architecture based on Personal Information Agents}
The \emph{architecture} of \cite{aaagrohStraub} is based on \emphtwo{personal information agents} associated with \emphtwo{spatio-temporally embedded social actors} (users, companies, SN-platforms etc.), which contextually decide upon the execution of another actor's query in relation to the asked actor's information space. The agents are able to answer these queries in a context sensitive way, using techniques from Context-Sensitive IR and their expertise on their own information spaces. Each actor maintains socio-semantic links to other spatio-temporally embedded actors in form of topic specific \emphtwo{expert-links}, thus implementing a special form of social spatio-temporal small world. Furthermore, each actor publishes a  selection of his / her expert-links and a set of spatio-temporally embedded \emphtwo{expertises}, summarizing content fields from the actor's information space (thus contributing to a special form of a semantic spatio-temporal small world). The spatio-temporal embedding of expert-links and expertises (`\emphtwo{knowledge flags}') follows the three step process discussed above. These knowledge flags are published in a decentralized \emphtwo{spatio-temporal Peer-to-Peer index}. If an actor issues a query which cannot be answered from his own information space, a social search is performed using the actor's expert links. If this search also fails, the spatial index is queried using the spatio-temporal embedding of the query, with the embedding following the three step process: e.g. if the query does not contain a spatio-temporal reference, the spatio-temporal reference of the questioner (see \autoref{humanir}) is used. The search delivers a number of knowledge flags which the questioner's agent then further evaluates by asking the related other agents. The system thus combines elements of social search (via expert-links), semantic search (local IR-systems) and spatio-temporal search (implying social and semantic contexts to a certain extent as explained above). 

Compared to e.g. Peer-to-Peer (P2P) IR systems, were e.g. an index is distributed over the passively protocol-executing peers in a P2P network (see e.g. \cite{tang2004hybrid} for a hybrid document- / index-distribution approach), and thus in most cases basically `merely' distributes a conventional IR system over a P2P network, this architecture uses the actor's agent's local IR systems to locally decide upon relevance. The agents are thus able to take into account the (e.g. social) context of the query and the querying agent / its user, thus being able to optimize contextual relevance and decide upon access  \cite{aaagrohBirnkammerer} to control information flows, ensure privacy or even employ information markets \cite{aaagrohBirnkammerer}). Furthermore, they are able to pro-actively keep their published knowledge flags up-to-date. 

The small world structure of the networks involved ensures that the expert-links, the comparatively coarse semantic mapping of the agent's information spaces in form of the expertises, and with the comparatively coarse implicit conserving of semantic and social contexts via spatio-temporal embedding is sufficient to deliver enough contextual seeds to reach enough competent agents which can then either employ their local IR systems to deliver contextually relevant items or use the private parts of their expert link list to further forward the query if the questioner's context is matching (e.g. if the corresponding user is a friend) resembling Human IR.  
\section{Study}
\subsection{Methodology}
%
Some elements of the architecture (such as the specially designed spatio-temporal P2P Quad-Tree) were evaluated using a dataset based on spatially referenced Wikipedia articles, demonstrating their practicability (see \cite{aaagrohStraubDonaubauerKoster},\cite{bbbBAKoster}). Despite not disposing of a full implementation and evaluation scenario involving the necessary large number of actors and sub-systems, another evaluation step that can be taken is to evaluate the suitability of the principles guiding the architecture's IR process inspired by Human IR, using social search, semantic search and spatio-temporal search and here especially the suitability of spatio-temporal embedding as a contextual bracket for this type of IR, implying social and semantic contexts to a certain extent as explained above. For this evaluation, a data-set is required that contains real association of users and information items as well as realistic locations of users and explicit spatio-temporal references of their information items, as well as a social network exhibiting characteristics of the expert-link network proposed in the architecture. The micro-blogging service Twitter \cite{twitter} with his network of followers, significant share of mobile usage and thus a large share of explicit spatial embeddings, and the free availability of the data is a suitable evaluation ground. We will now discuss some results of this evaluation.
%
%
%
%
%
\subsection{Dataset}
A dataset from Twitter was downloaded in June and July 2010, using the Twitter API \cite{twitter}. The Tweets and Re-Tweets which were non-English (which was decided using the approach described in \cite{cavnar1994n}, employing an ML classifier using language specific n-gram statistics) were discarded. The remaining (Re-)Tweets were Porter-stemmed \cite{porter1980algorithm} and stop-words were removed. Of the Re-Tweets, only the additional content without `re-citing' the original Tweet was regarded. An undirected social network between the users was induced by establishing an edge if at least one @Reply or @Mention \cite{twitter} (roughly corresponding to a direct message) was exchanged between the respective users. Of this social network, the largest connected component was chosen, and the rest of the users and their Tweets and Re-Tweets discarded. We downloaded the complete information from 43973129 Tweets and Re-Tweets, of which 9725514 were explicitly geo-coded. 3323803 of these geo-coded entities were associated with the largest connected component of our social network and finally considered. Of the 6887632 users in the dataset, 670271 were explicitly geo-coded and 160690 of these belonged to the largest component of the social network that we considered. 

Users were spatially embedded via the geo-location of their last available explicitly geo-located (Re-)Tweet.  
(Re-)Tweets not explicitly spatially embedded (via geo-coordinates) were embedded with a simple geo-parsing approach, analyzing the strings denoting the location and subsequently using the MetaCarta geo-coding service \cite{metacarta}. 
If this process failed, the geo-location of the Wikipedia articles corresponding to the tags of the respective (Re-)Tweet, were used, using the Wikapidia API (see previous section). If that fails, the location of the authoring user was used. Locations were subjected to very small (uniform distribution in [-0.1,0.1] decimal degrees) random deviations to avoid mapping many entities to the exact same location which would result in overcrowding peers with respect to the Quad-Tree based spatio-temporal index 
which was used in the evaluation environment for the experiments. 
\subsection{Interrelations between Spatio \-Temporal, Social and Semantic Contexts}
The \emph{social network}'s mean average path length was 6.92 (a random graph with the same number of nodes, which was computed with the help of the JUNG framework \cite{jung} yielded a value of 8.96), and the average clustering coefficient \cite{watts1998collective} of the social network was 0.046 (corresponding random graph: 0.000014). We see that although the average clustering coefficient on SN platforms is usually higher by a factor of $>4$ (e.g. \cite{wilson2009user}report an average 0.164 for their early 2009 crawl of several sub-networks of Facebook with an overall number of nodes of $\approx 10^6$). The numbers indicate that the present network can still be regarded as having small world properties. 
\begin{figure*}[htpb]
\centering
\subfigure[\label{degree}Degree distribution of social network]{\includegraphics[height=0.20\textheight]{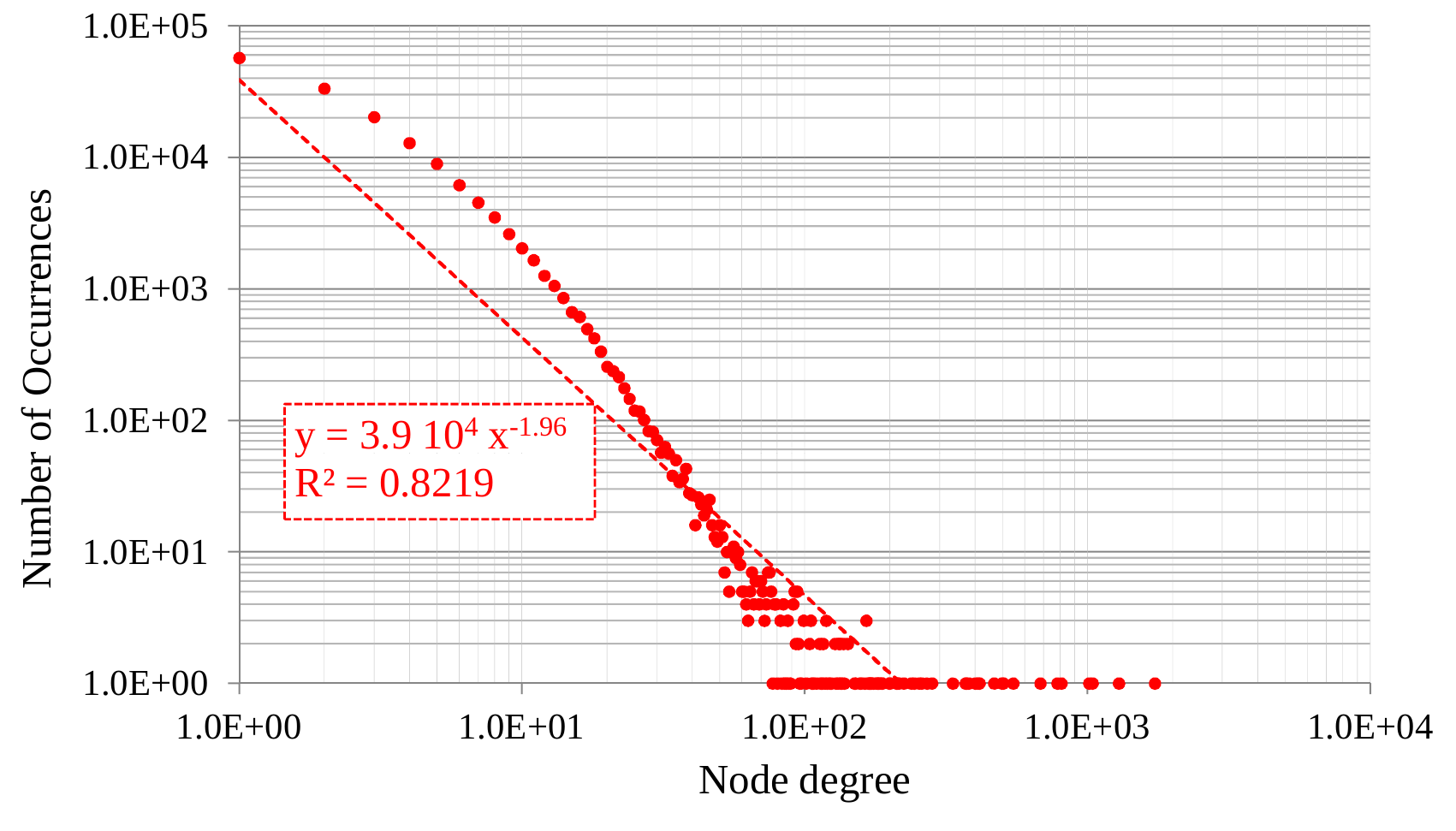}}
\subfigure[\label{nrtweets}Distribution of number of (Re-)Tweets per user]{\includegraphics[height=0.20\textheight]{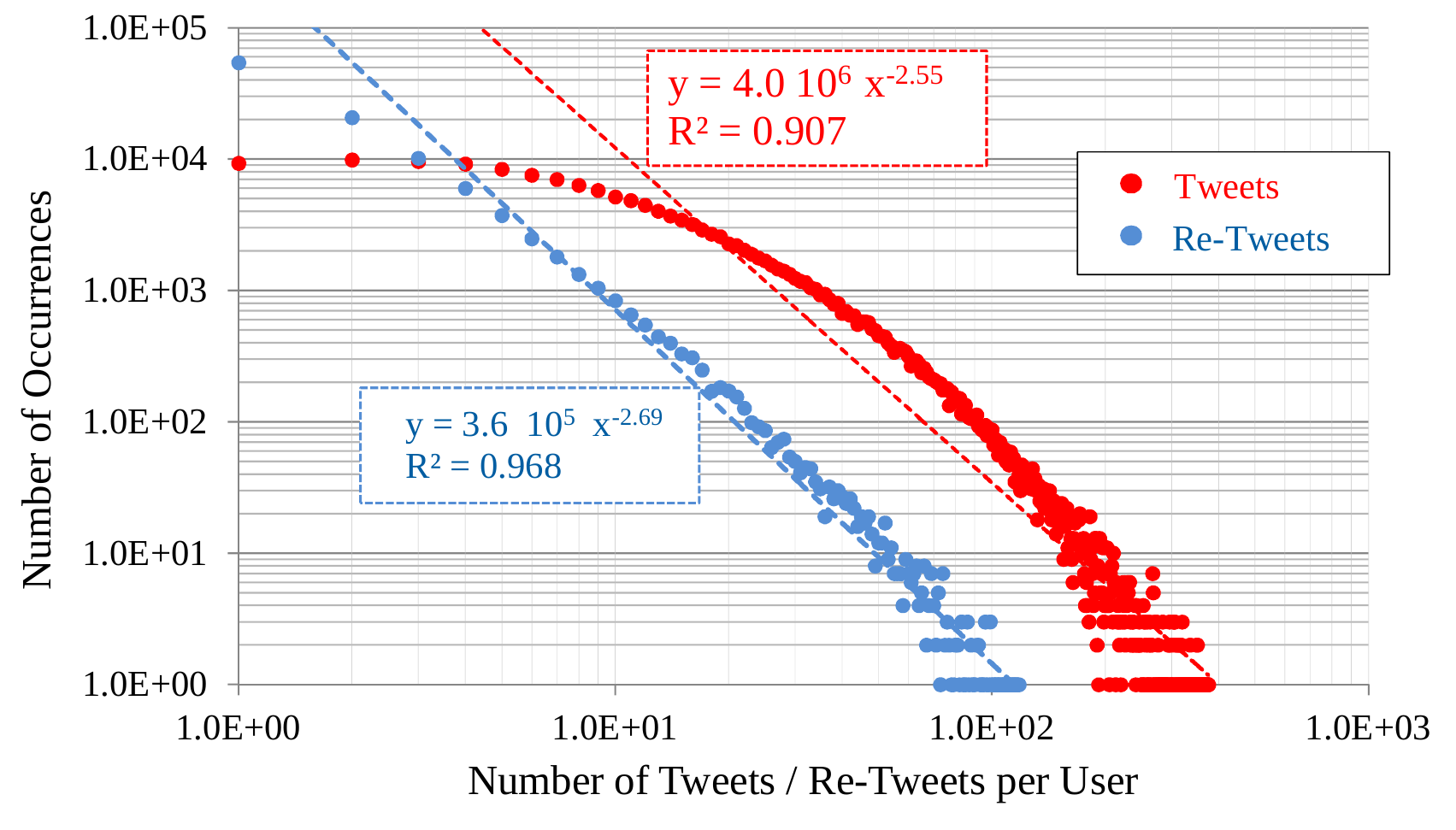}}
\subfigure[\label{distance}Distribution of distance between adjacent nodes in the social network]{\includegraphics[height=0.20\textheight]{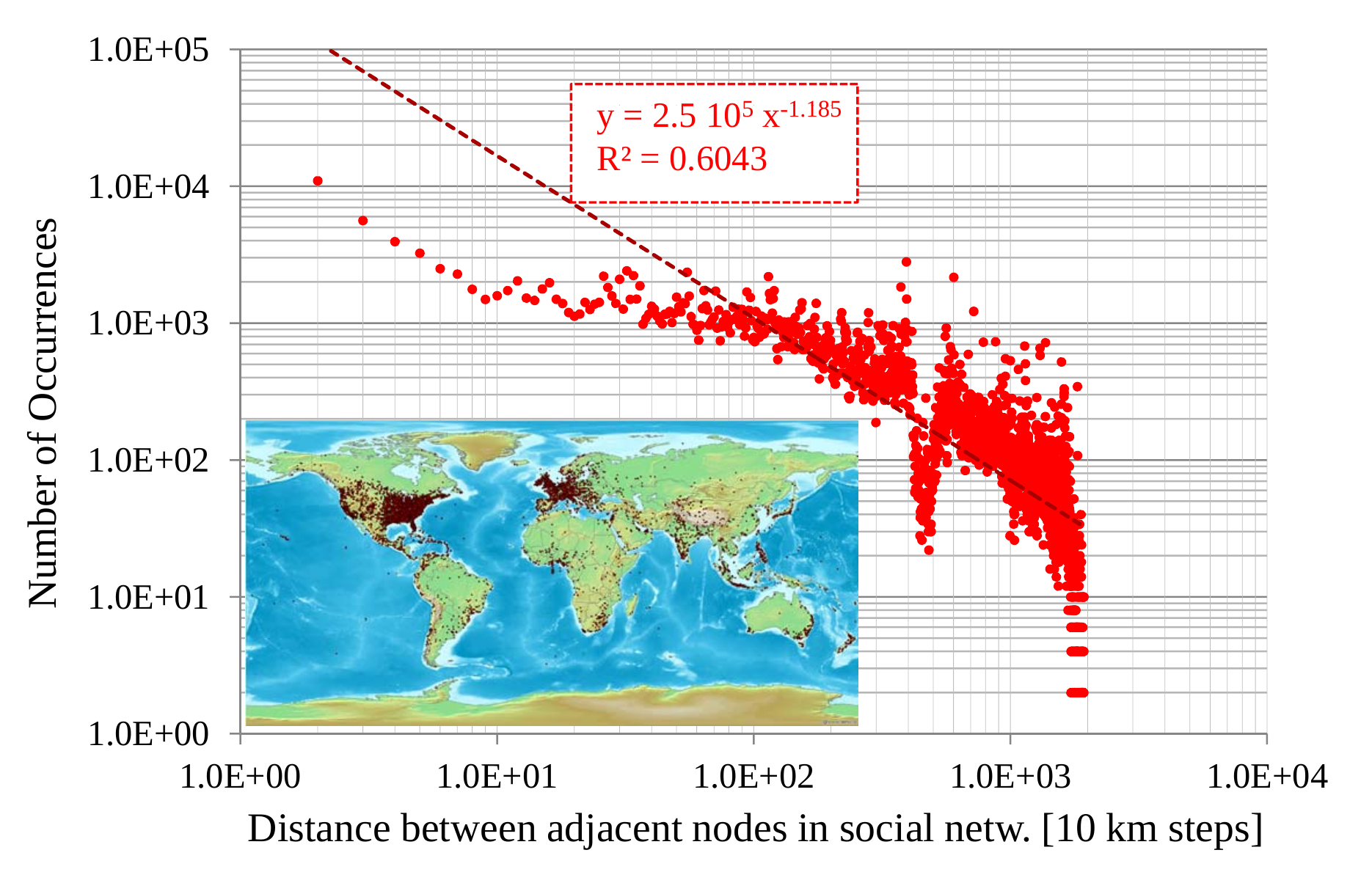}}
\subfigure[\label{spatialdistsemanticsim}Correlation between spatial distance and semantic similarity of information spaces]{\includegraphics[height=0.20\textheight]{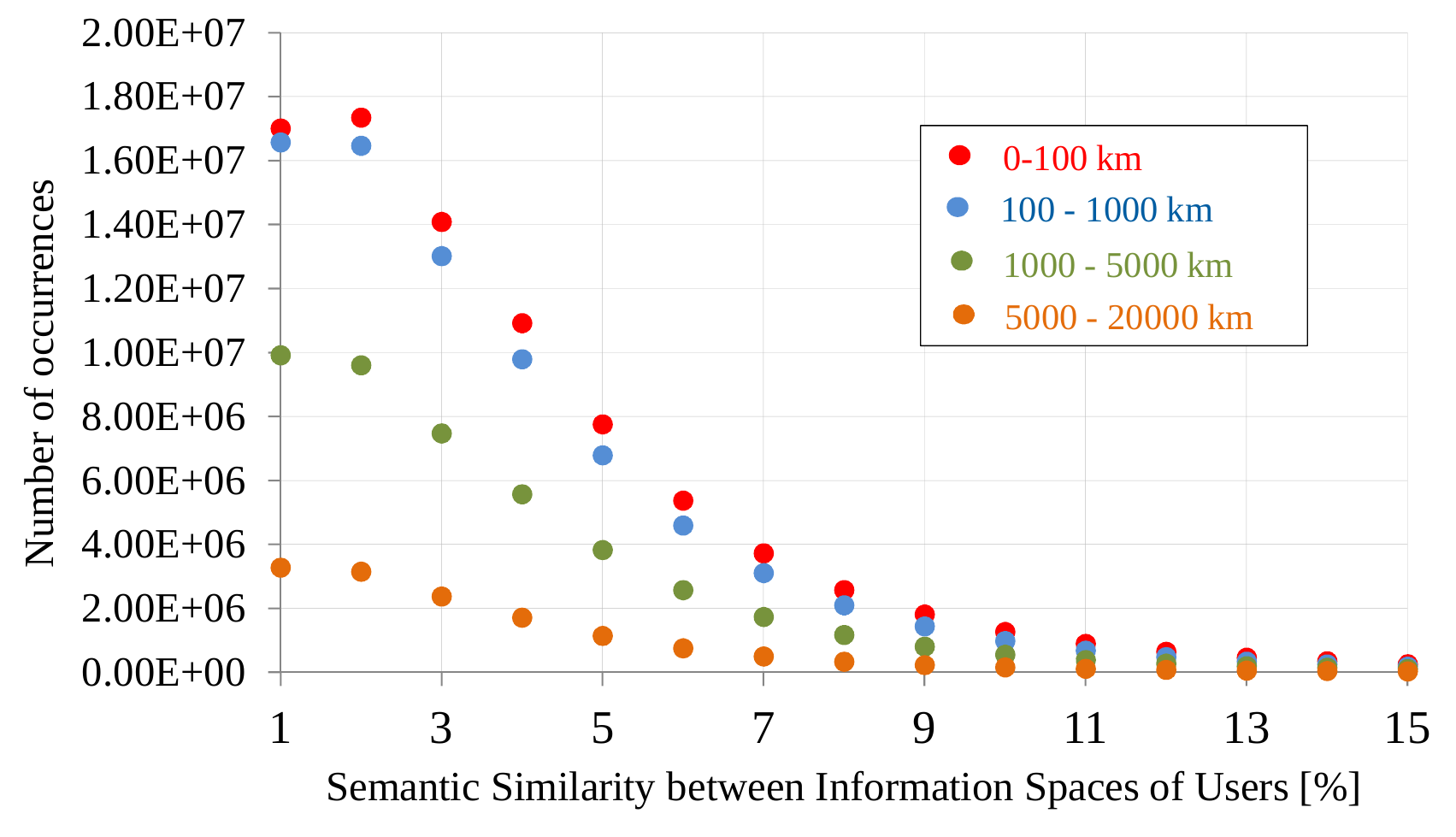}}
\subfigure[\label{networkdistsemanticsim}Correlation between network distance and semantic similarity of information spaces]{\includegraphics[height=0.20\textheight]{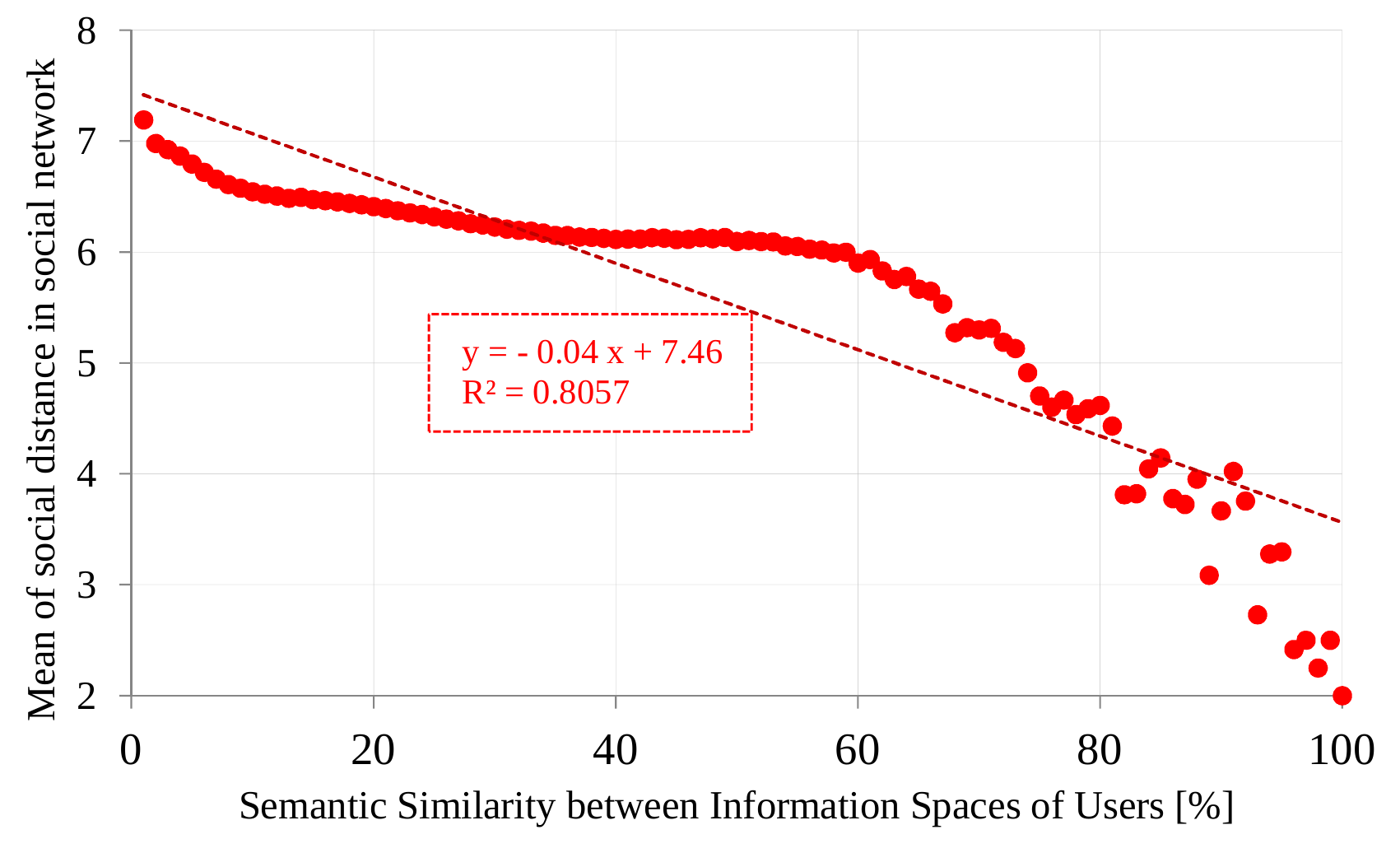}}
\subfigure[\label{geographicsemanticsim}Correlation between geographic and semantic similarity of information spaces]{\includegraphics[height=0.20\textheight]{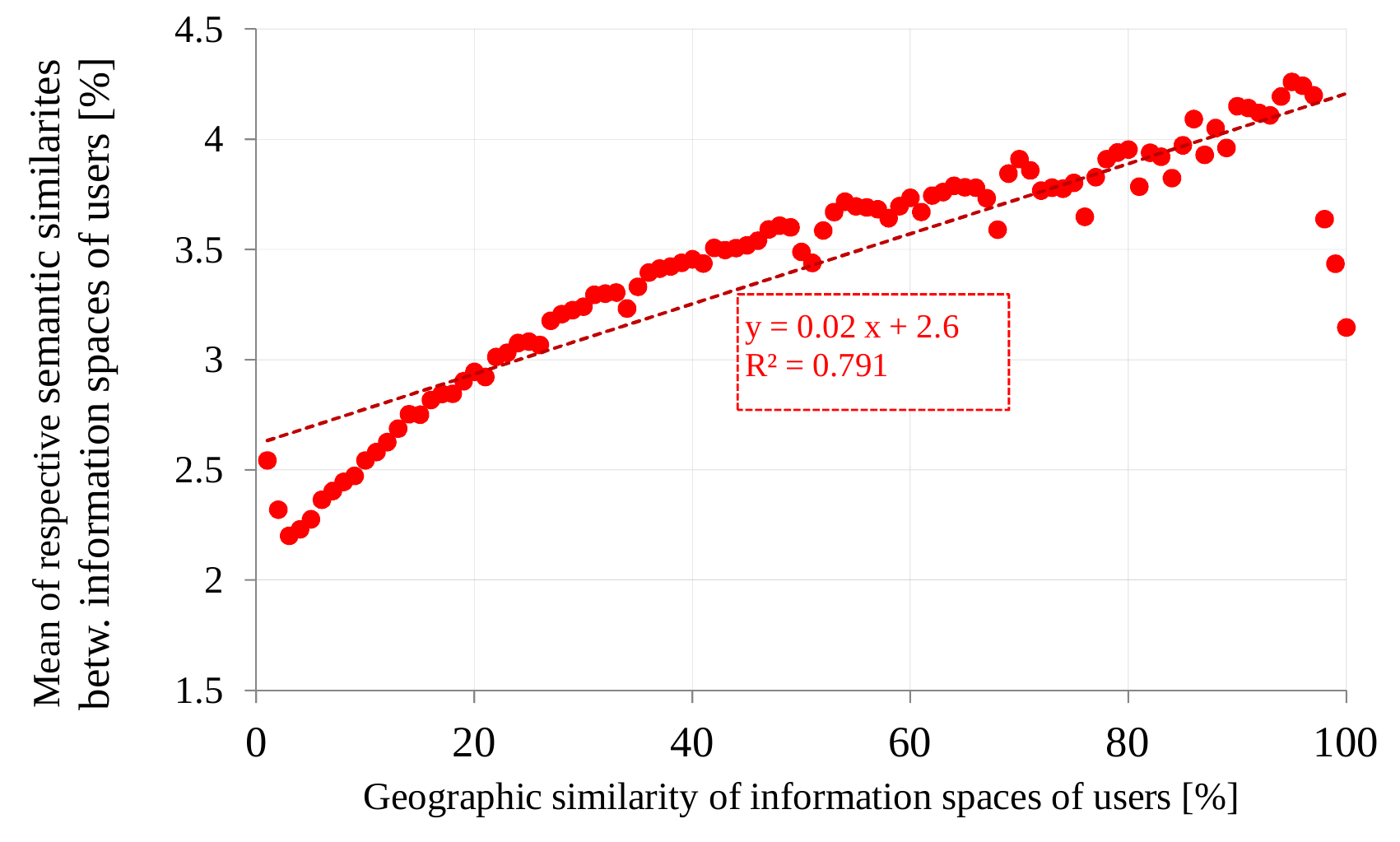}}
\subfigure[\label{semanticsimspatialdist}Correlation between geographic similarity of information spaces and spatial distance of corresponding users]{\includegraphics[height=0.20\textheight]{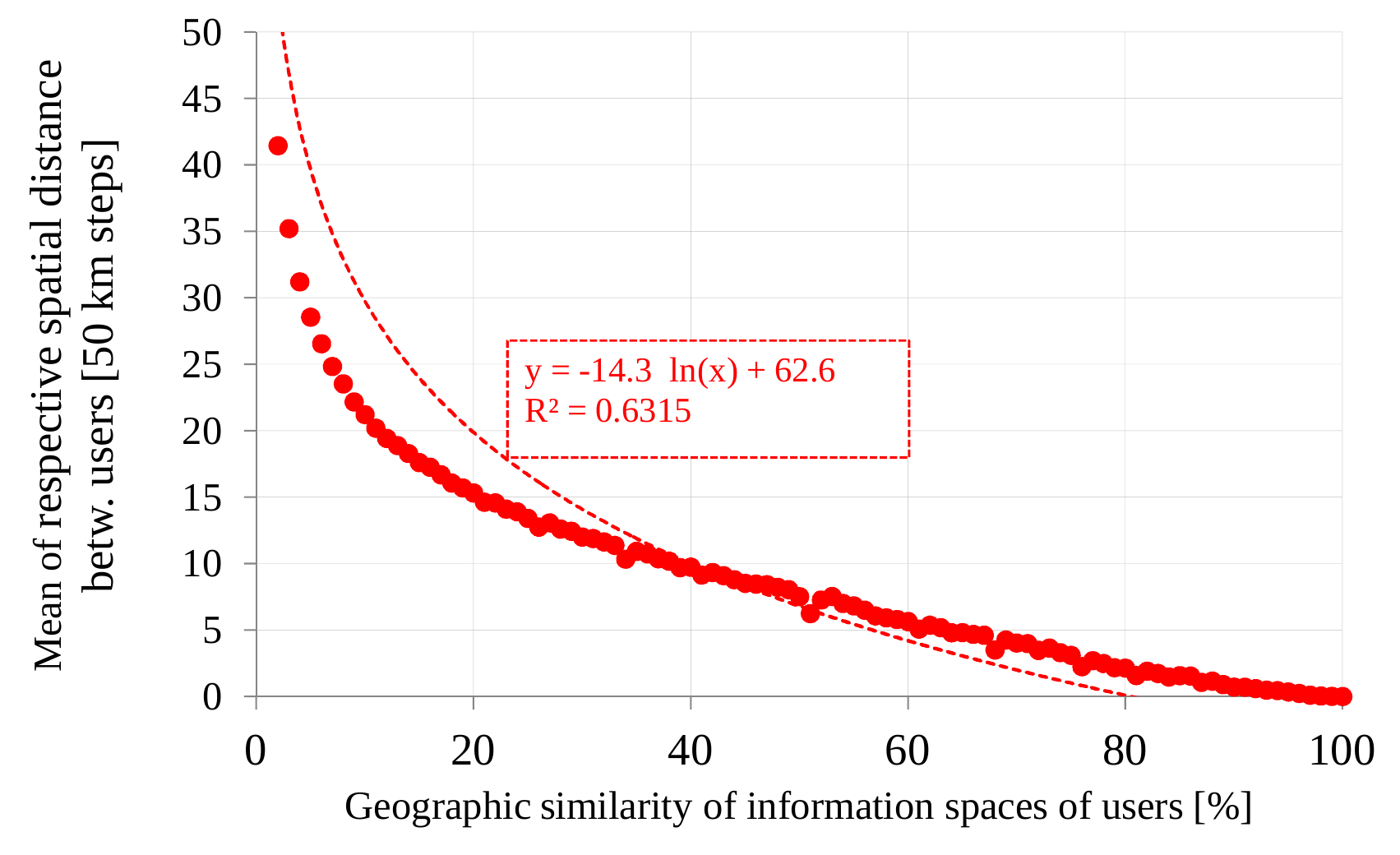}}
\subfigure[\label{socialsemanticsimilarity}Correlation between semantic similarity of information spaces social similarity (via friend sets)]{\includegraphics[height=0.20\textheight]{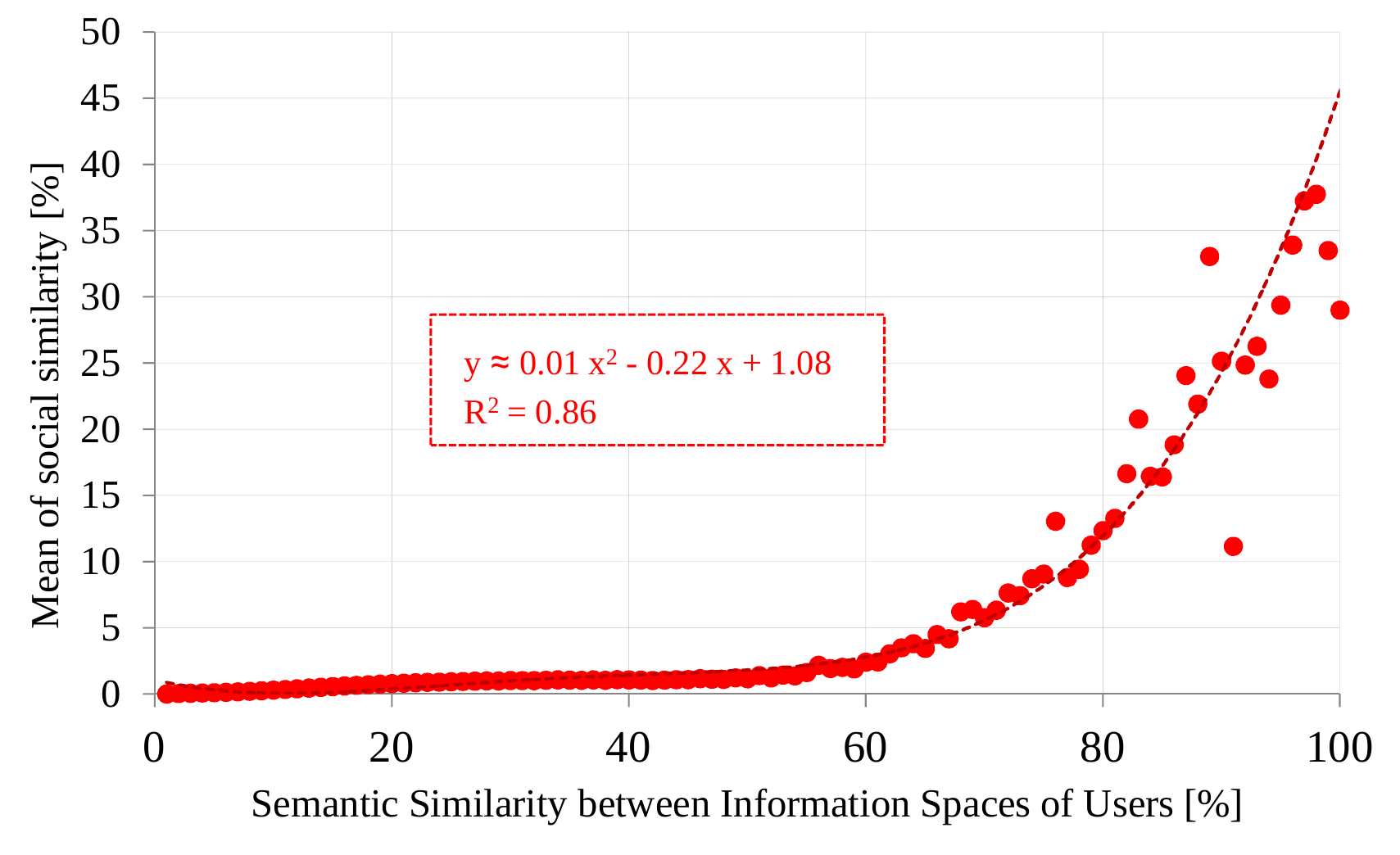}}
\caption{General properties of the dataset and mutual implication of social, semantic and spatial closeness using different measures (compare discussion in the text) \cite{grohhabil}. 
Wherever a curve fit is provided (e.g. a power law, linear or logarithmic function), standard regression \cite{pruscha2005statistisches} is used where $R^2 = 1 - \sum_i(y_i-f(x_i,\beta))^2 / \sum_i(y_i-\bar{y})^2 $ is the coefficient of determination \cite{grohhabil}.}
\label{straubBenni}
\end{figure*}
%

\emph{\autoref{straubBenni}} shows statistical properties of the dataset and correlation effects that support the mutual implication of social, semantic and spatial closeness which represents a basis for the proposed IR architecture. \emph{Sub-Figure \ref{degree}} shows the degree distribution of social network which roughly follows a power law. This fact and the deviations from the exact power law distribution coincide with the findings in \cite{newman2003structure} \cite{libennowell2005geographic}. Together with the previously discussed values for the mean average path length and clustering coefficient shows that the social network of actors in the data-set can indeed be assumed to be a realistic small world social network. 

\emph{Sub-Figure \ref{nrtweets}} shows a distribution of the number of Tweets and Re-Tweets per user which, in our experiment simulate the information spaces of the users. While the Re-Tweet distribution follows a power law, the distribution of the number of Tweets shows deviations from the power-law distribution, while the $R^2$-value of fitting an exponential function $y(x)=ae^{-bx}$ is significantly lower, supporting that a pure exponential fit is less appropriate. Functions of the type $y(x)=\beta x^{-\alpha}+ae^{-bx}$, which induce an exponential cutoff of the power-law's long tail, qualitatively show a better congruence with the distribution and intuitively correspond to the reasonable assumption that extremely large sizes of information spaces of users in SN and MSN platforms are very rare. 

\emph{Sub-Figure \ref{distance}} shows the distribution of spatial (geodesic) distance between adjacent nodes (actors with a direct social relation) in the social network. Equivalence classes of geodesic distances are determined in steps of 10 km. Due to the spherical topology of earth's surface (with a maximum circumference of roughly 40000 km at the Equator), the maximum class of spatial distances encompasses all geodesic distances between 19990 km and 20000km. As reasonably expected, the distribution shows two users with a smaller spatial distance have a higher probability of being socially connected, where the distribution roughly follows a power law. This confirms other study's results, such as \cite{libennowell2005geographic}  
and supports the assumption that social closeness and spatial closeness mutually imply each other to a certain extent. As the diagram depicted in the left corner of the diagram shows, the geographic distribution of the users concentrates on the densely populated areas of North America and Europe. The dip of the curve around $\approx 5000km$ may be explained by the relative geometric dimensions of the Atlantic ocean and the North American and European continent. 

\emph{Sub-Figure \ref{spatialdistsemanticsim}} shows the correlation between the spatial distance of pairs of users (this time counted in classes of 50 km steps) and the semantic similarity of their information spaces (counted in equivalence classes of 1 \%). The semantic similarity of information spaces was computed as the Tanimoto coefficient \cite{lipkus1999proof} of the multi-set of term-frequency vectors of the respective sets of information items. Other alternatives would have e.g. been to use Rocchio centroids \cite{joachims1997probabilistic}. As an implementation, we used Lucene \cite{lucene}. 
Of a matrix containing the absolute frequency of occurrences for a combination of a geodesic distance class  and  class of semantic similarity of information spaces we computed the average absolute frequencies for the four new equivalence classes $[0, 100 \mathrm{km}]$,  $[100, 1000 \mathrm{km}]$,  $[1000, 5000 \mathrm{km}]$, and  $[5000, 20000 \mathrm{km}]$. The four qualitatively Gaussian curves show that for larger distances the semantic similarity of the information spaces of the users is smaller than for smaller distances. This supports the connection between semantic relatedness and geographic relatedness. Qualitatively similar results have been obtained by \cite{hecht2008geosr} although the measures used were different.   
 
\emph{Sub-Figure \ref{networkdistsemanticsim}} depicts a correlation between the semantic similarity of information spaces of users (computed as in sub-figure \ref{spatialdistsemanticsim}) and their average path distance in the social network.  (Technically: of a matrix containing the absolute frequency of occurrences for a combination of a class of semantic similarities between $[x,x+1]$\% and a path distance in the social network, we computed for each class of semantic similarities between $[x,x+1]$\% the average over all path distances between 0 and 25). The result shows that the more similar the information spaces the smaller is the average social distance between the respective users. This supports the correlation between social closeness and semantic closeness.  

\emph{Sub-Figure \ref{geographicsemanticsim}} shows a correlation between the geographic similarity of information spaces of users and their semantic similarity. While semantic similarity was computed in the same way as in 
\ref{networkdistsemanticsim} and \ref{spatialdistsemanticsim}, the geographic similarity of information spaces of users was computed in the following way: In order to compute a spatial relevance density for the information space of a user,
a point-like spatial reference $\mu=(\mu_1,\mu_2)$ of an information item was transformed into a Gaussian density contribution $\mathcal{N}(\mu,\sigma)(x)$ with diagonal sigma corresponding to a 500 km circle, cut off  at $|x-\mu| = 500$ km with the help of ArcGis \cite{arcgis}. All contributions (which properly respected the spherical geometry of earth's surface) were added to yield a user $u_i$'s spatial relevance density $\rho_i(x)$. The geographic similarity $\mathrm{sim}(u_i,u_j)$ of the information spaces of two users $u_i$ and $u_j$ was computed via a Jaccard-like measure: 
\begin{equation}
	\mathrm{sim}(u_i,u_j)=\int \mathrm{d}^2x \frac{\min (\rho_i(x),\rho_j(x))}{\max(\int \mathrm{d}^2x \rho_i(x), \int \mathrm{d}^2x \rho_j(x))  }
\end{equation}
Although most of the information spaces had a similarity of 0 (this large contribution was left out of the diagram) the values show a trend that the closer the geographic similarity of information spaces, the larger the semantic similarity. 
Although the slope of this trend is rather small, this finding supports the correlation between geographic reference of information spaces and their semantic similarity. 

Relating this geographic similarity of information spaces to the spatial geodesic distance between users as shown in \emph{sub-figure \ref{geographicsemanticsim}}, yields a logarithmic trend supporting the reasonable connection that spatial closeness of users also implies similarity in the spatial references of their information spaces. 

\emph{Sub-Figure \ref{socialsemanticsimilarity}} relates the social similarity between users computed as the Jaccard-index of the sets of friends of two users and the respective average semantic similarity of information spaces (where the semantic similarity of information spaces is computed as in sub-figures \ref{spatialdistsemanticsim}, \ref{networkdistsemanticsim}, and \ref{geographicsemanticsim}. We see a power law relating the two quantities: the more socially similar two users are, the more similar are their friend-sets and vice versa. This supports the connection between social and semantic contexts. 

These preliminary results are an excellent ground for future research, investigating the connections between social, spatio-temporal and semantic contexts. 
\FloatBarrier
\subsection{Information Retrieval Experiments}
The results just discussed show that the dataset can be viewed as a dataset \emph{realistically} including and relating social, spatial and semantic elements. They \emph{support the basic findings} of \autoref{infoneeds} and \autoref{humanir} and the \emph{grounds} for the IR approach discussed in \autoref{irapproach}. In order to \emph{evaluate the basic suitability of these connections for IR}, IR experiments were conducted with the data-set. 

As \emph{queries}, Tweets were used. In the absence of real user assessments of relevance to be used as \emph{ground truth} for the experiments, two \emph{implicit assessments of relevance} were used as ground truths:  
As a \emph{first} assessment of relevance, the \emph{Re-Tweets} of the query Tweet were regarded as relevant. This assessment of relevance is intended to represent relevance with respect of the \emph{conscious information need} of users. As a \emph{second} assessment of relevance, all Tweets and Re-Tweets of users \emph{following} (see \cite{twitter}) the author of the query Tweet were regarded as relevant. This assessment of relevance is intended to represent relevance with respect to the \emph{unconscious information needs} of users containing the contextual seeds discussed in \autoref{infoneeds} and \autoref{humanir}. 

In order to compare semantic search, social search and spatial search (excluding temporal aspects for reasons of simplicity) as a contextual bracket implicitly relating social and semantic contexts, \emph{seven types of retrieval processes} were tested on the data-set. For each type of retrieval, the 50 best results (according to the IR model of the respective type) are retrieved and analyzed with the \emph{first (I)} and \emph{second (II)} 'ground truth' assessment of relevance by computing the usual confusion matrix (TP, FP, TN, and FN) and from that precision P and recall R \cite{manning2008introduction} 
If less than 50 items could be retrieved, either the missing ones are padded with random items from the respective pre-filtering (e.g. geographic or social) (\emph{variant A}) before computing the measures to ensure comparability, or the measures are computed as is (\emph{variant B}). 

\emph{Type 1} $[$Sem$]$: semantic search (standard IR): Use Lucene \cite{lucene} to compute a global IR index (over all information items of the dataset) and decide upon the 50 best matches to the query Tweet using Lucene's ranking. 

\emph{Type 2} $[$Soc$]$: social search (social pre-filtering and subsequent semantic filtering): Retrieve all information items authored by friends and friends of friends of the query Tweet's author, compute a local IR index on these items and  decide upon the 50 best matches to the query Tweet using the local index. This type of search is roughly associated with the expert-link-based type of social search with subsequent evaluation using a local IR system in the architecture. 

\emph{Type 3} $[$Geo$]$: geographic search (geographic pre-filtering and subsequent semantic filtering): Using our implementation of our variant of distributed Quad-Tree and an octagonal query geometry centered around the query Tweet's spatial point reference of 'radius' between 500 km and 20 km depending on the depth of the tree in this region (corresponding to the density of information items), the spatially matching items were retrieved. On this set of items the semantically 50 best were determined as in the case of social search. This type of search is roughly associated with the spatio-temporal search of the architecture on Expertises with a subsequent employment of local IR. 

\emph{Type 4} $[$Soc$\cup$Geo$]$: social-geographic search $\cup$ (using the union $X\cup Y$ of the results of geographic $X$ and social pre-filtering $Y$ and subsequent semantic filtering with Lucene as in type 2 and 3). This type is roughly associated with the spatio-temporal search of the architecture on all knowledge flags (Expertises and Expert-Links) with subsequent local IR. 

\emph{Type 5} $[$Soc$\cap$Geo$]$: social-geographic search $\cap$ (using the intersection $X\cap Y$ of the results of geographic $X$ and social pre-filtering $Y$ and subsequent semantic filtering). This type of search is performed for reference purposes.

\emph{Type 6} $[$RndGeo$]$: random pre-filtering geographic (randomly select as many items from the dataset as a geographic pre-filtering would deliver and perform subsequent semantic filtering). This type of search is performed for reference purposes to further investigate the impact of geographic pre-filtering and thus the role of spatial context as a contextual bracket. 

\emph{Type 7} $[$RndSoc$]$: random pre-filtering social (randomly select as many items from the dataset as a social pre-filtering would deliver and perform subsequent semantic filtering). This type of search is performed for reference purposes to further investigate the impact of social pre-filtering. 

\begin{figure}[htpb]
\centering
\subfigure[\label{precisionia}Precision I]{\includegraphics[height=0.39\columnwidth]{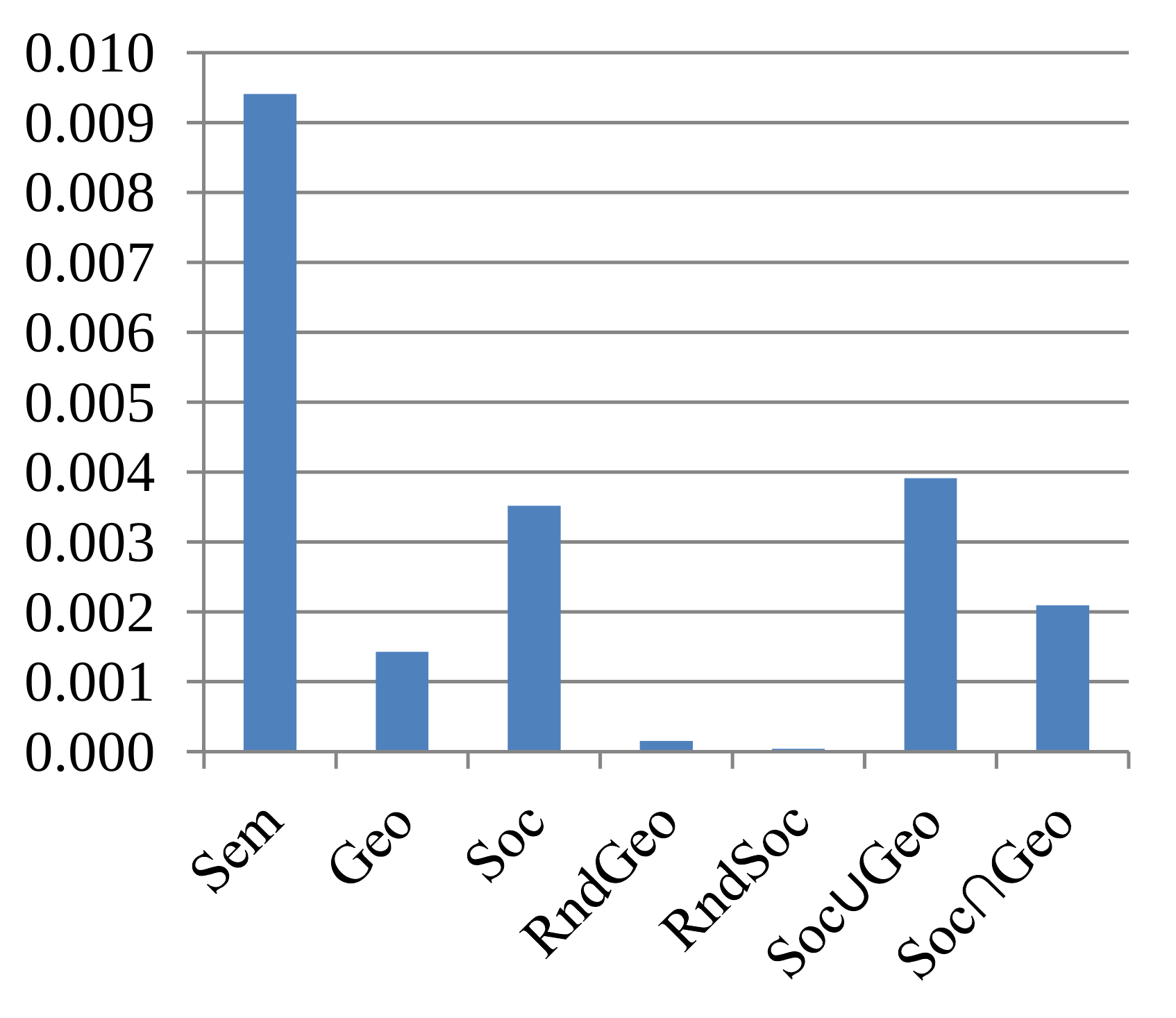}}
\subfigure[\label{precisioniia}Precision II]{\includegraphics[height=0.39\columnwidth]{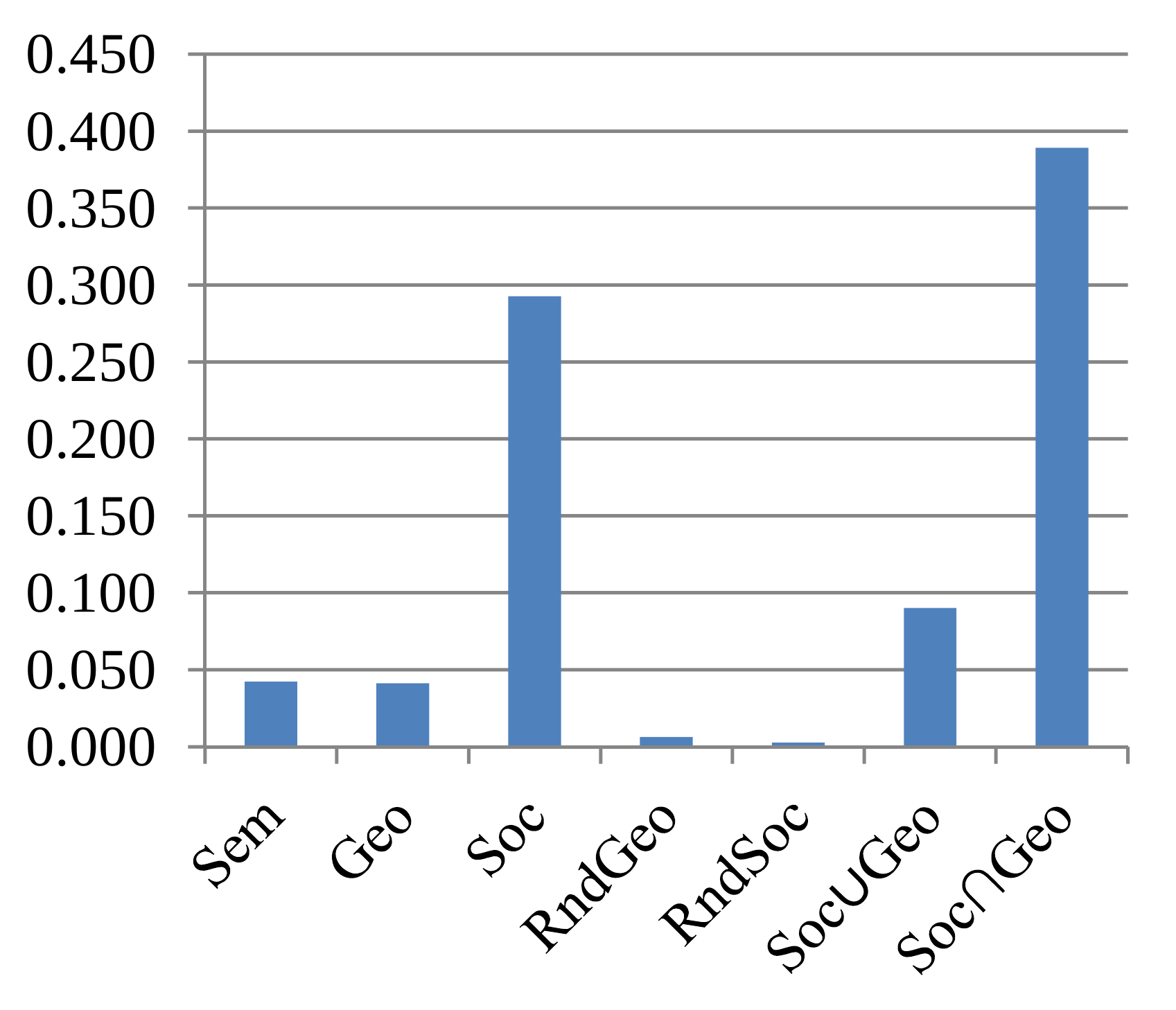}}
\subfigure[\label{recallia}Recall I]{\includegraphics[height=0.39\columnwidth]{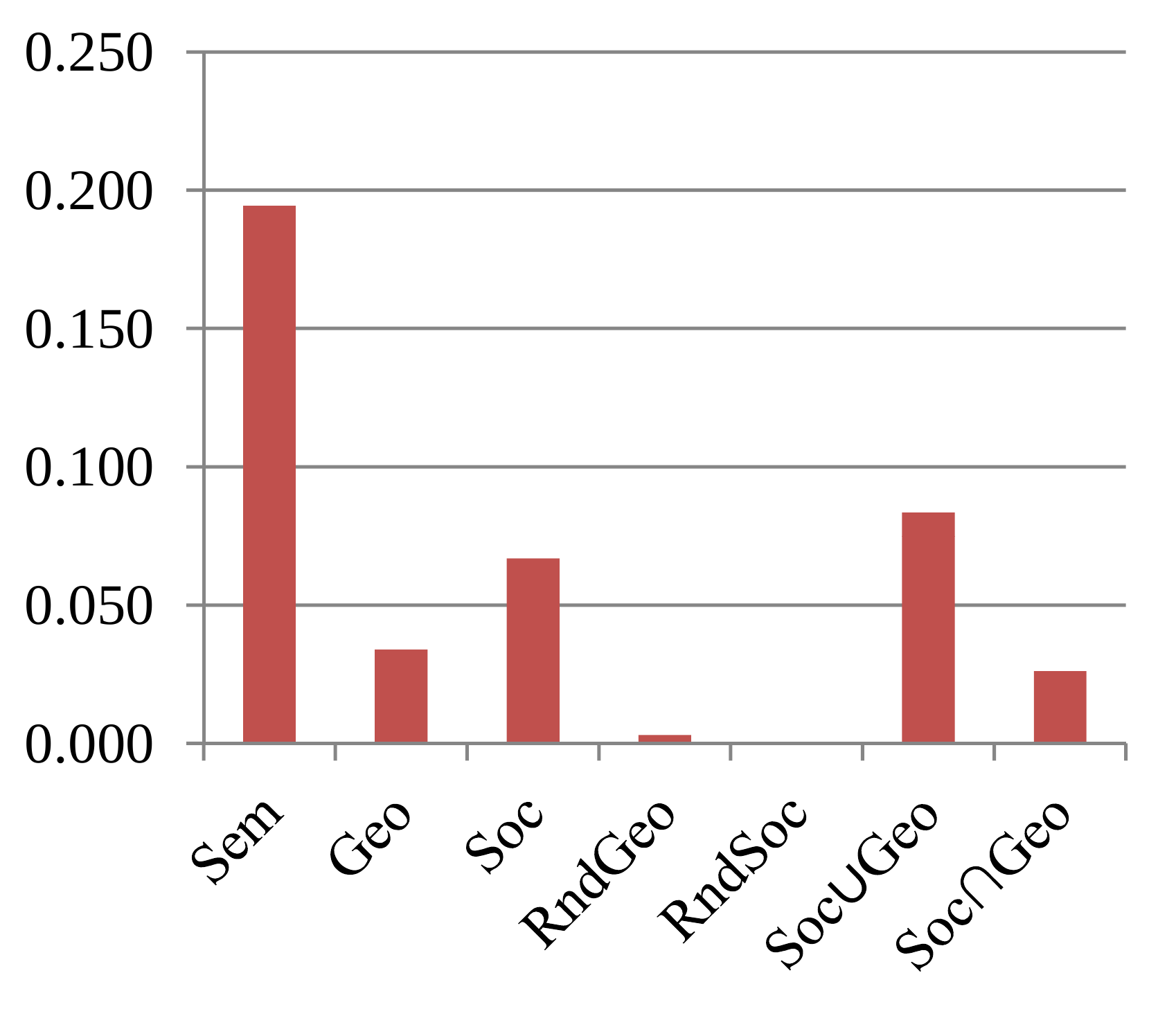}}
\subfigure[\label{recalliia}Recall II]{\includegraphics[height=0.39\columnwidth]{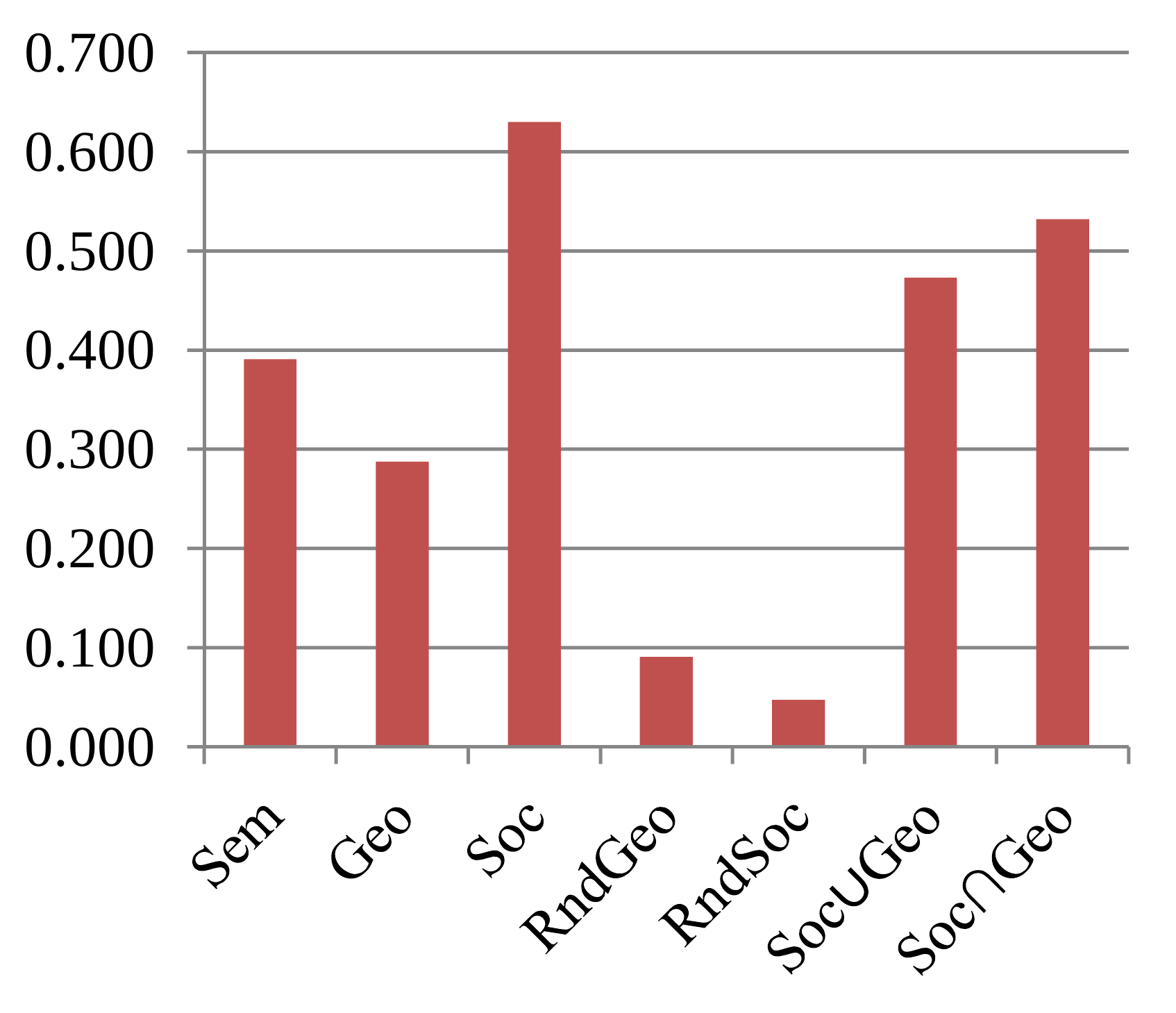}}
\caption{Precision and Recall, variant A \cite{grohhabil}}
\label{straubBenni2A}
\end{figure}
\autoref{straubBenni2A} shows the precision and recall values of variant A. The while for I, the first way of ground truth relevance assessment, the conventional purely semantic search performs best by far (in precision as well as recall), social search is most successful for II, the second way of ground truth relevance assessment and geographic search is still comparable to semantic search. If the assumption that II corresponds to contributing to satisfying unconscious information needs via contextual seeds is indeed substantial, this result supports the proposed IR approach. In view of the role of spatial context as a context bracket implying semantic context to a certain degree, the comparison of the performance of geographic search (Geo) compared to random pre-filtering geographic (RndGeo) shows that indeed, Geo is significantly better than RndGeo. In other words, while Sem may use the whole set of information items to choose the 50 best (via the global index), Geo must choose from the considerably smaller set resulting from geographic pre-filtering and still delivers acceptable relative performance compared to a random pre-filtering. Indexing the whole set of information items may not be desirable for SN and MSN environments due to privacy considerations. Because of the connections between geographic closeness and social closeness, we can thus, in a realistic SN and MSN setting, expect that Geo may effectively draw from a locally richer set of relevant items and thus deliver even better overall performance than Sem.   

\begin{figure}[htpb]
\centering
\subfigure[\label{precisionib}Precision I]{\includegraphics[height=0.39\columnwidth]{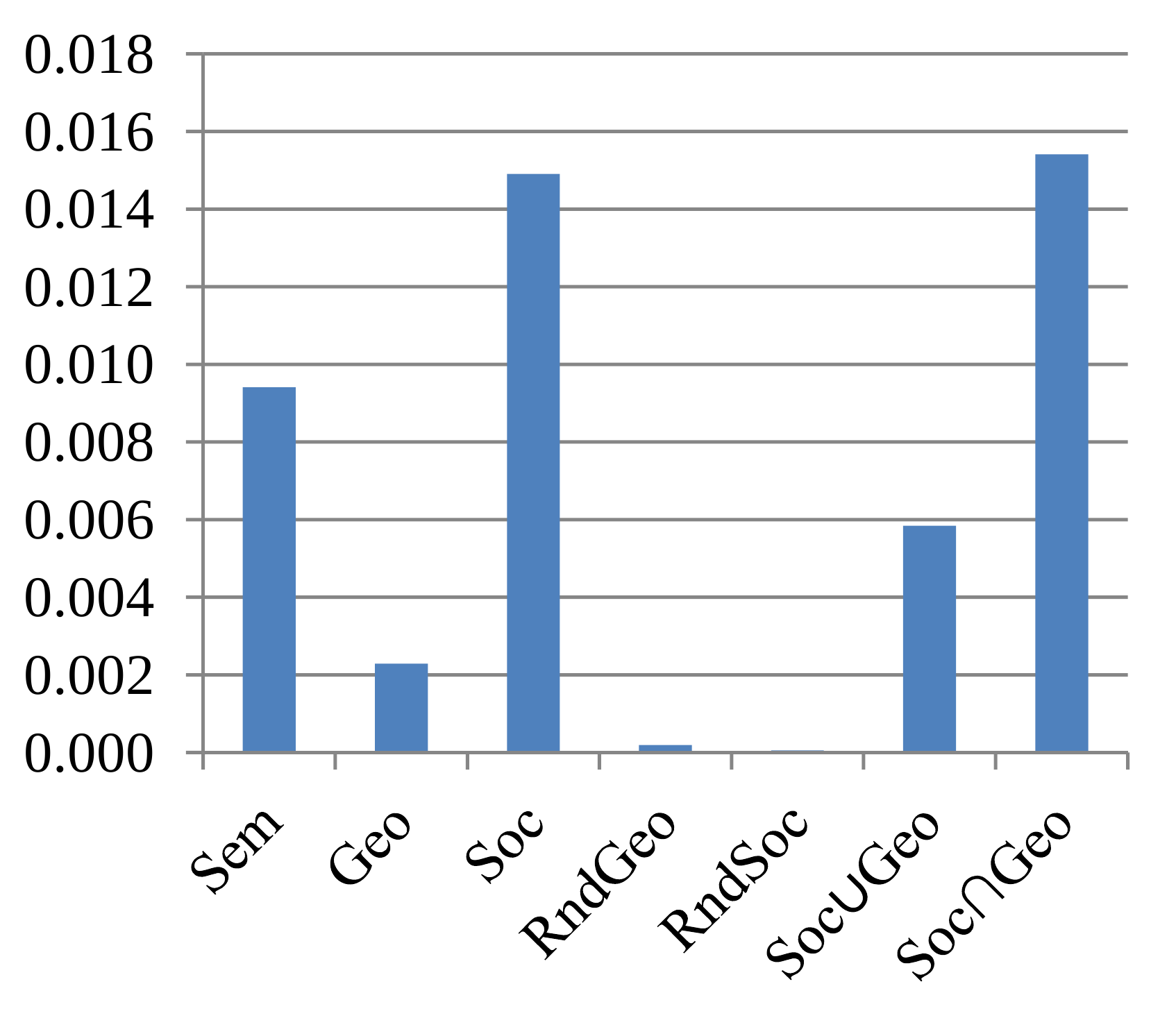}}
\subfigure[\label{precisioniib}Precision II]{\includegraphics[height=0.39\columnwidth]{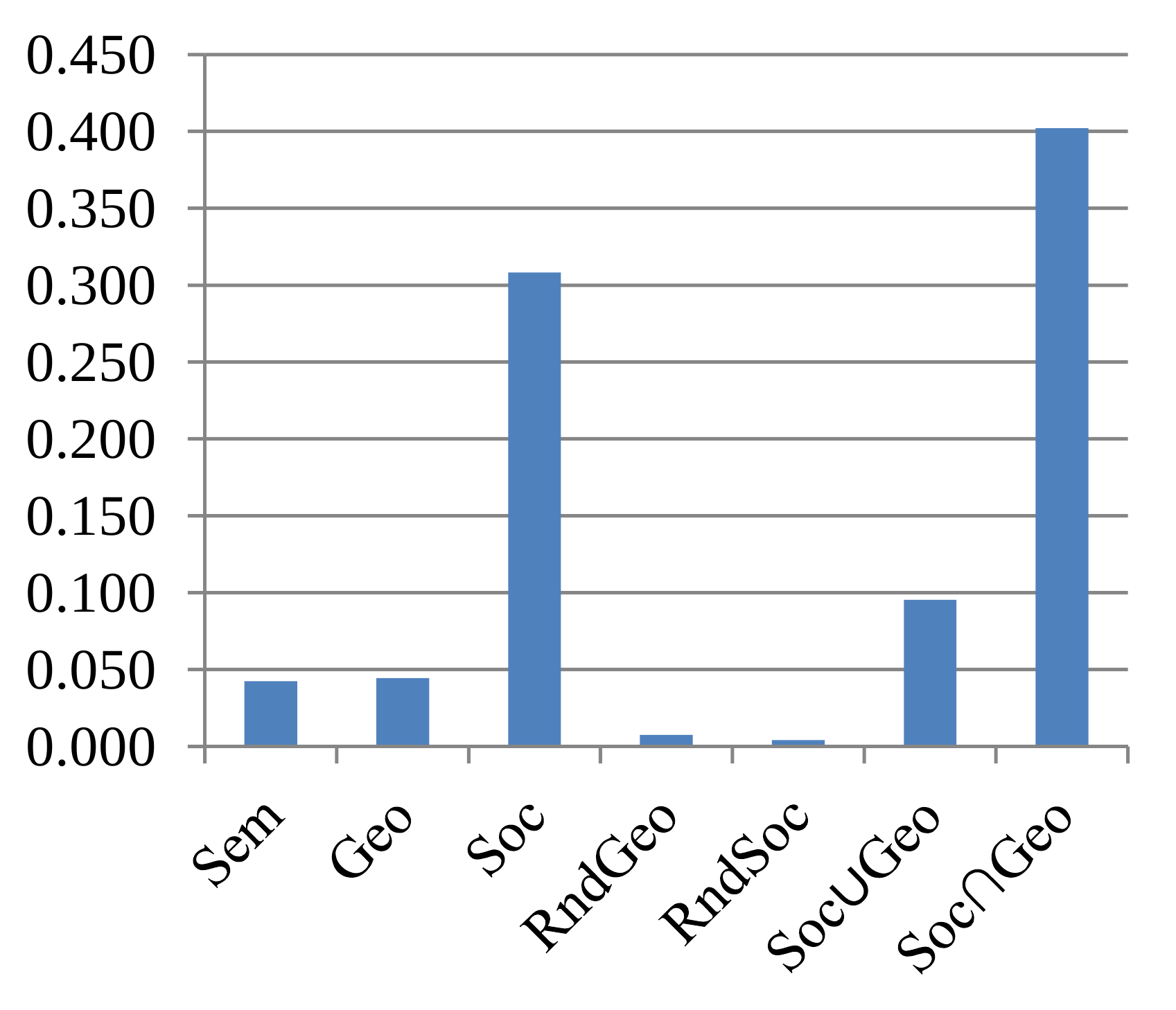}}
\subfigure[\label{recallib}Recall I]{\includegraphics[height=0.39\columnwidth]{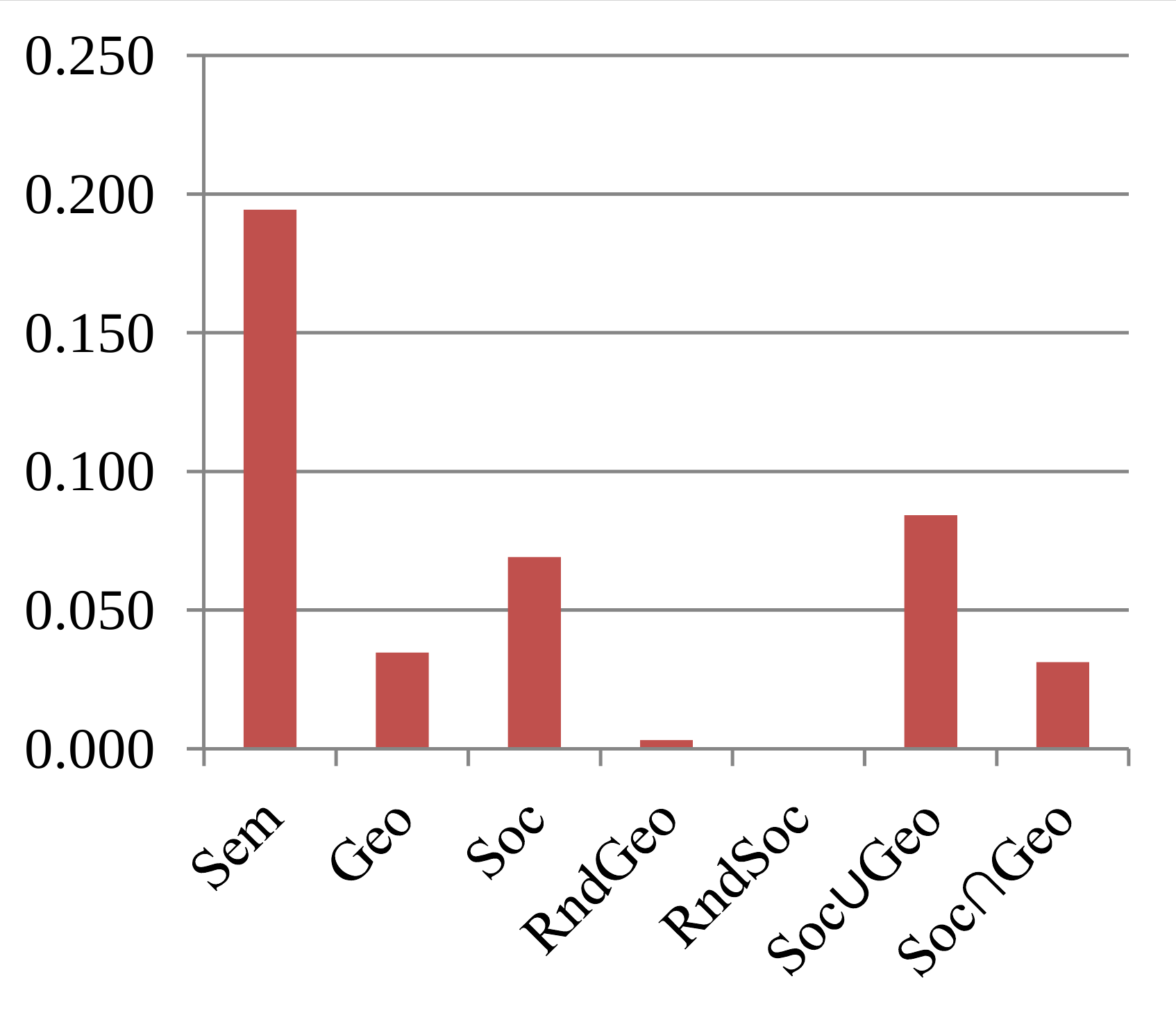}}
\subfigure[\label{recalliib}Recall II]{\includegraphics[height=0.39\columnwidth]{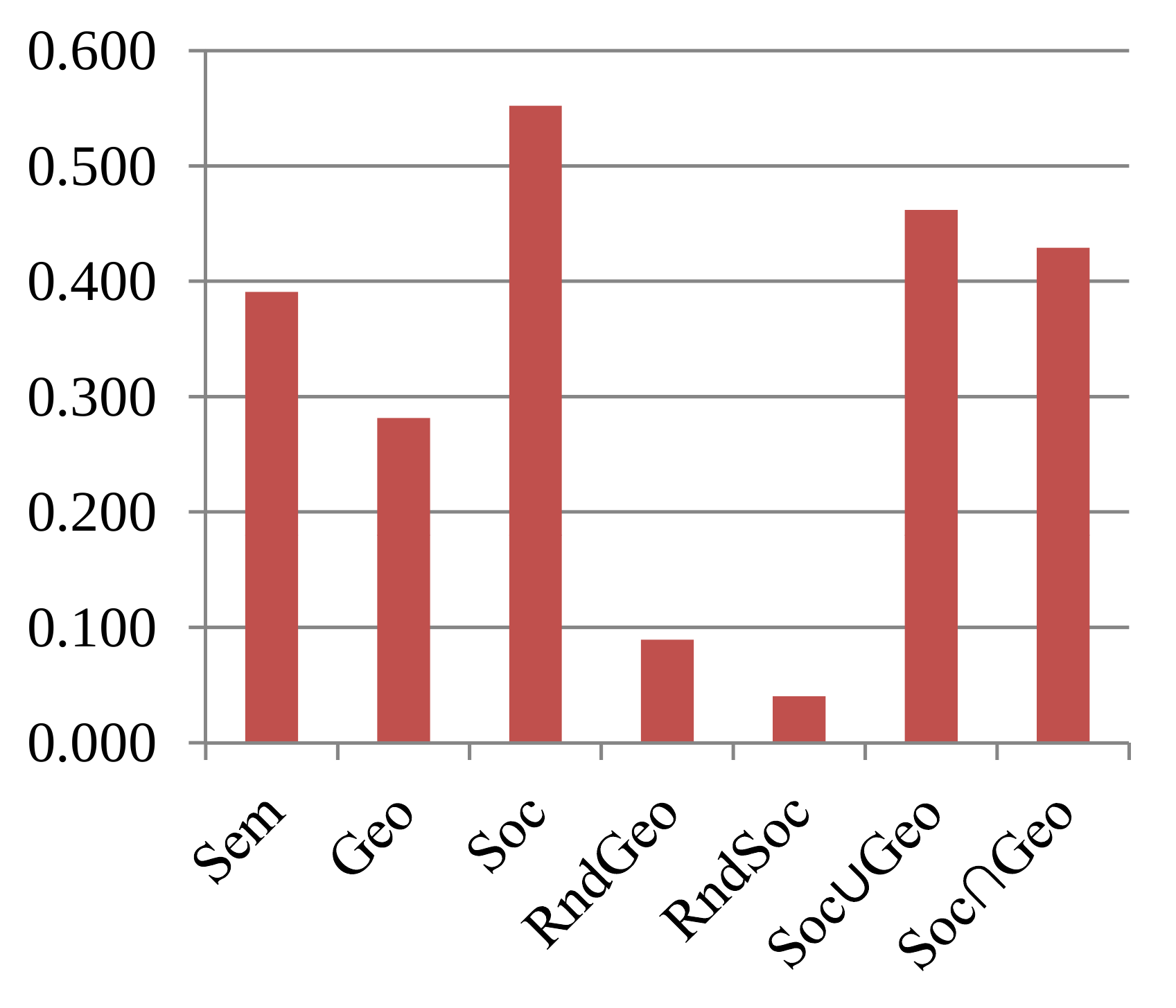}}
\caption{Precision and Recall, variant B \cite{grohhabil}}
\label{straubBenni2B}
\end{figure}

\autoref{straubBenni2A} shows the precision and recall values of variant B, where a, due to the restrictive pre-filtering, insufficient number of retrieved items is not padded by random items (which induces a pessimistic evaluation for the contextual search variants). Here, as a consequence, social search is best also for assessment I with respect to precision. 

\vspace{4ex}
\section{Conclusion and Opportunities \\for Future Research}
Our \emph{overall results} may be interpreted as giving \emph{support} to exploiting the concept of Spatio-Temporal Small Worlds and the underlying correlations between semantic, spatio-temporal, and social contexts for alternative IR, akin to Human IR in (Decentralized) Social Networking.  

However, the evaluation environment may still not take advantage of several of the benefits of the architecture (such as the power of local agent IR systems). Thus, one might expect that the approach is indeed able to deliver useful contextual seeds especially in view of unconscious information needs and thus is a new alternative IR concept for SN and MSN environments. 

Nevertheless, the introduced study is only a starting point for a large body of \emph{future work} on connecting social, semantic and spatio-temporal contexts for new and useful forms of IR.   

As has been mentioned above, a full implementation and real world evaluation of the architecture would be the next step following the usual Design Science methodology \cite{hevner2004design}. A special focus has to be put on evaluating the usefulness of the results obtained by the suggested alternative IR methods in terms of the extended notions of information need discussed above. Suitable concepts of extended versions of precision and recall will have to be constructed for the respective evaluations. Furthermore, more variants of combining spatial, social and semantic retrieval criteria need to be evaluated in relation to the individual and social short term and long term context of the querying user.  
\bibliographystyle{plain}
\bibliography{georgBibFile}
\end{document}